\documentclass[a4paper,10pt]{article}
\usepackage[utf8]{inputenc}
\usepackage{jheppubMod}
\usepackage{amsmath,amsfonts,amssymb,graphicx,MnSymbol,subfigure}
\usepackage{hyperref} 
\usepackage{tikz}
\usetikzlibrary{shapes.geometric, arrows}
\usepackage[toc,page]{appendix}
\usepackage{adjustbox}

\title{Surveying the theory space of pion dark matter}
    \abstract{We use a holographic model to survey the space of strongly coupled SU($\nc$) gauge dynamics with  QCD-like chiral symmetry breaking pattern for quarks in the fundamental representation. We systematically identify the light degrees of freedom ($\rho, \sigma$ and $\pi$ mesons) that would make up the dark sector as a function of $\nf, \nc$ and a common quark mass scale. We identify seven distinct effective theories that are of interest to explore and make a first summary of the expected dark matter phenomenology. Amongst our results we conclude that QCD-like models where the low energy theory is described purely in terms of pions struggle to generate a large enough $M_\pi/f_\pi$ value so these theories will need extra relic density generation mechanisms to be viable. The largest space of models (with an intermediate quark mass) have both the $\sigma$ and $\rho$ lying below $2 M_\pi$ and are largely unexplored in the literature so far. Regions with just one of $\rho$ or $\sigma$ light are possible in constrained parameter regions. These states aid the pion relic density generation as needed for valid theories. The $\sigma$ always lies above the $\pi$ mass becoming degenerate with the $\pi$ in the extreme walking limit. }

\author[a]{Anja Alfano}
\author[a]{Nick Evans,}
\author[b]{Suchita Kulkarni,}
\author[c]{Werner Porod,}
\affiliation[a]{Department of Physics and Astronomy, University of Southampton,  Southampton, SO21 2LJ, UK}
\affiliation[b]{Institute of Physics, NAWI Graz, University of Graz, Universit\"atsplatz 5, A-8010 Graz, Austria}
\affiliation[c]{
Institute for Theoretical Physics and Astrophysics, Julius-Maximilians-Universit at Wurzburg, 
97074 Wurzburg, Germany}
\emailAdd{a.alfano@soton.ac.uk}
\emailAdd{evans@soton.ac.uk}
\emailAdd{suchita.kulkarni@uni-graz.at}
\emailAdd{werner.porod@uni-wuerzburg.de}
\date{March 2024}

\newcommand{\newc}{\newcommand}
\newcommand{\nc}{{N_c}}                
\newcommand{\nf}{{N_f}}                

\newcommand{\mpi}{{M_\pi}}
\newcommand{\mpisq}{{M^2_\pi}}
\newcommand{\mpicube}{{M^3_\pi}}
\newcommand{\fpi}{{f_\pi}}

\newc{\nev}[1]{\textcolor{blue}{(Nick: #1)}}
\newc{\sk}[1]{\textcolor{red}{(Suchita: #1)}}
\newc{\wrp}[1]{\textcolor{cyan}{(Werner: #1)}}

\newc{\UV}{{\mathchoice{}{}{\scriptscriptstyle}{}UV}}
\newc{\IR}{{\mathchoice{}{}{\scriptscriptstyle}{} IR}}
\begin{document}
\maketitle


\section{Introduction}

An interesting possibility is that dark matter is the pseudo-Nambu Goldstone bosons (pNGB) (or  pions) of a dark sector strongly coupled gauge theory. Such theories provide a new relic density mechanism by means of the $3 \pi\to 2\pi$ Weiss-Zumino-Witten process. They also generate a large enough self-interaction cross-section that might  explain the core versus cusp problem in astrophysical data. These theories are collectively known as strongly-interacting massive particle (SIMP) theories~\cite{Hochberg:2014kqa}.

The original SIMP models have a spectrum where all bound states ($\rho, \sigma$ etc) are heavier than twice the $\pi$ mass and so decay in the strongly coupled sector leaving purely a theory of pions in the infra-red (IR). These models though need a large value of $\mpi/\fpi$ to reconcile relic density by $3 \pi\to 2\pi$ processes and self-interaction cross-section constraints. Depending on the theory details, the SIMP models may also feature additional relic density mechanisms, where heavier states such as $\rho$ or $\sigma$ play an important role~\cite{Pomper:2024otb,Bernreuther:2019pfb,Bernreuther:2023kcg,Choi:2018iit,Appelquist:2024koa}. These extra annihilation channels alleviate the tensions between relic density and self-interaction limits in the basic SIMP model. In this paper we seek to more systematically analyse the space of possible infra-red effective models. We will concentrate here on SU($\nc$) gauge theories with mass-degenerate fermions in the fundamental representation and a chiral symmetry breaking pattern SU($\nf)_L \times$ SU($\nf)_R \rightarrow SU(\nf)_V$.  Throughout this paper we will use the mass of the $\rho$ meson, $M_\rho$, at $m_Q=0$ as a measure of the strong coupling scale - we denote it $M^0_\rho$. Our analysis concentrates on the $\pi$ which are light as pNGB of the chiral symmetry breaking, the scalar singlet $\sigma$ since it potentially becomes light as a pNGB of conformal symmetry breaking in theories with a walking gauge coupling, and the $\rho$ as the lightest vector state which can be phenomenologically relevant. Our aim is here is to establish the properties of light states  that will be important for possible dark matter scenarios, rather than carrying out full cosmological analyses or attempting strongly-interacting dark matter model building.

The main obstruction to simply enumerating SIMP scenarios is their inherent non-perturbative nature. The low energy properties of the theory, which emerge from the strongly coupled theories, such as the pion masses and decay constants can not be directly  calculated using any known perturbative methods. This means, the low energy parameters of the theory, although controlled by high energy inputs, need to be determined using methods beyond ordinary perturbation theory. Ideally one would use lattice gauge theory, a first principle approach, but the computational time and expense is huge. Even in the presence of lattice data, practical computation of all phenomenological implications needs to be done using low energy effective field theories, which need to be constructed. There is current work on determining the form of the effective theories particularly at large number of flavours ($\nf$) to colours ($\nc$) ratio, taking the theory in the walking regime~\cite{DelDebbio:2021xwu,Zwicky:2023krx}. 

We will make use of a holographic model to explore these theories~\cite{Alho:2013dka,Erdmenger:2020flu}. The model works well for the lightest meson sector of QCD (at the 20\% level quantitatively~\cite{Clemens:2017udk}) and incorporates the dynamics through an input running coupling for a theory as a function of $\nc,\nf$. For a fixed number of colours ($\nc$), the model incorporates the approach to the conformal window at some $N^c_f$ critical value of the number of flavours. The model displays a light $\sigma$ ``dilaton" in the walking regime - the more conformal symmetry is restored near the chiral symmetry breaking scale the lower the $\sigma$ mass is \cite{Evans:2013vca}. The model has simple $\nf, \nc$ scalings for the decay constants of the theory compatible with the UV theory. The model also includes arbitrary quark masses ($m_Q$). Whilst the model is not a first principles computation it does allow us to explore possible behaviours in the space of $\nf, \nc, m_Q$ and identify possible low energy effective theories and  estimate the tuning needed to achieve them. The holographic model can also be used to describe quarks in higher dimension representations \cite{Erdmenger:2020lvq,Erdmenger:2020flu,Alfano:2024aek,Alfano:2025dch} but here we concentrate on the fundamental representation - the main change resulting from using higher dimension representations would be that the decay constants grow with the dimension of the representation which would not aid raising $M_\pi/f_\pi$.

\begin{figure}[ht]
    \centering
    \includegraphics[width=0.47\linewidth]{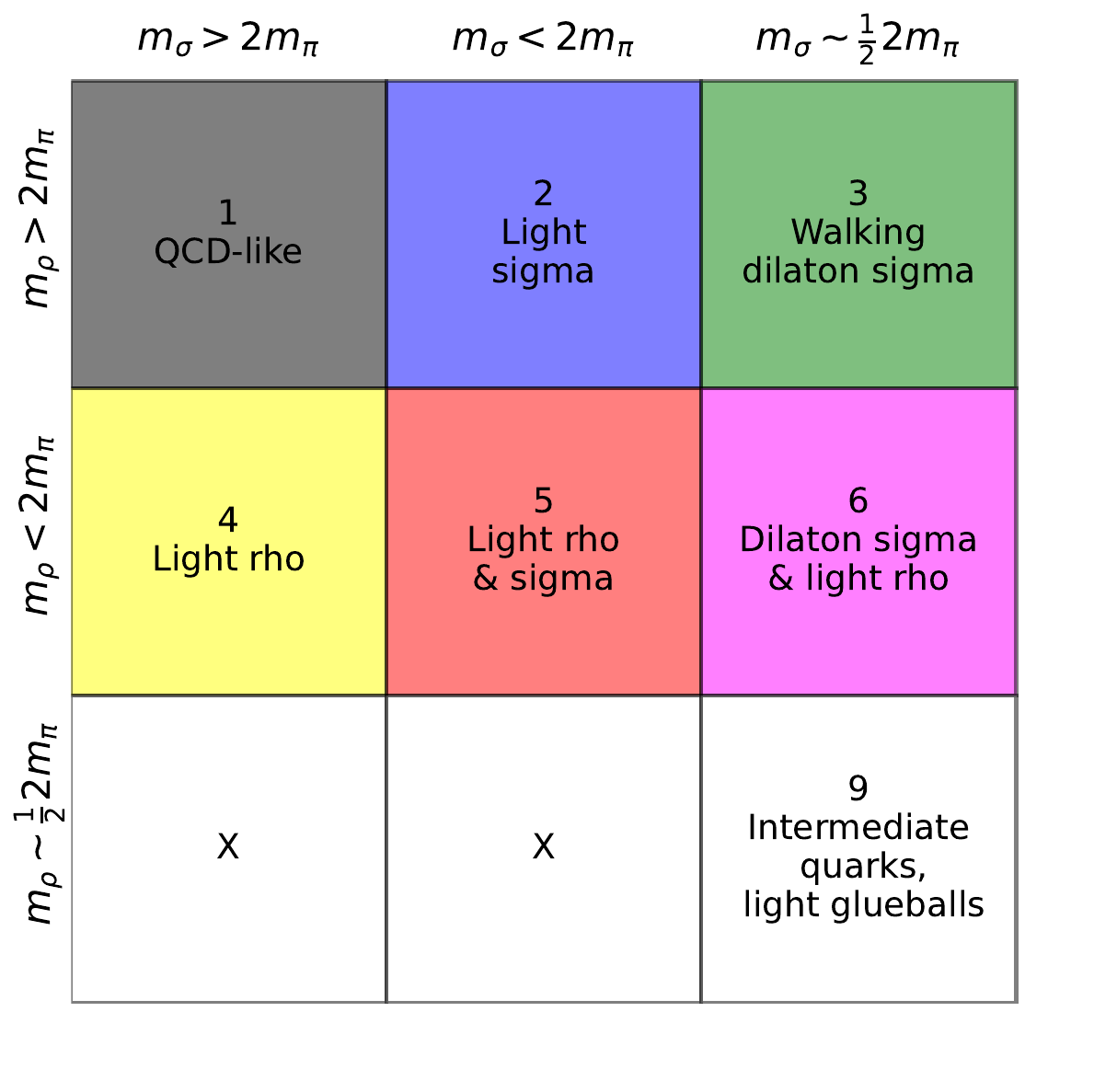}
    \includegraphics[width=.45\linewidth]{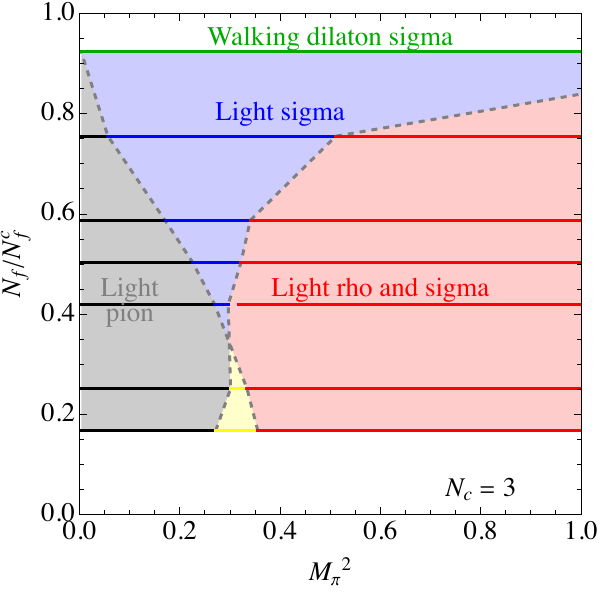}
    \caption{Left panel: The regions of parameter space denoted by their low energy spectrum. Right panel: the classification of the $\nc=3$ theory, using the colours from the left in the $N_F/N_F^c$ versus $m_Q$ plane. The horizontal lines are for integer $N_F$ where we compute. Regions 1-5 are present at these $M_\pi$ (in units of $M_\rho^0$).   Regions 6 and 9 lie to the right of the right panel where the gauge theories are transiting to weak coupling and holography is less reliable. }
    \label{fig:boxes_1}
\end{figure}

In total we identify seven different possible IR regimes which are summarized in fig.~\ref{fig:boxes_1} (left panel). In fig.~\ref{fig:boxes_1} (right panel) we show the results of the holographic model revealing where in the $\nf/\nf^c$ versus $\mpisq$ plane each phase is likely to be found. As we will see this is largely $\nc$ independent. Note that $\mpisq$ is a measure of the quark mass, at least at small $m_Q$. 

We will begin by presenting the holographic model (sec.~\ref{sec:model}) that can describe the strongly coupled phases (Regions 1-6,9 in fig.~\ref{fig:boxes_1} (left panel)). In sec.~\ref{sec:mass_spec} we will determine and present fits for the $\sigma, \pi$ and $\rho$ masses and decay constants across the $\nf, \mpisq$ plane at $\nc=3$. We will discuss the simple $\nc$ scaling the model holds to allow extrapolation of these results to higher $\nc$. These results then lead to the phase space depicted in fig.~\ref{fig:boxes_1}(right panel). 
 Regions 1--6 have an IR sector made from some or all of the fields $\pi,\rho,\sigma$. The regions 7 and 8 in fig.~\ref{fig:boxes_1} can not be realised since we know of no mechanism to parametrically lower the $\rho$ mass. Region 9 is expected to describe the meson sector of gauge theories where $m_Q$ lies near or above the strong coupling scale of the theory. If the quark mass is very large one expects the gauge dynamics to 
be weakly coupled at the mass scale. The meson masses will lie close to $2 m_Q$. However, we also expect at these large $m_Q$ that the gauge dynamics will survive below $m_Q$ going on to become strong in the deeper IR. It will generate light glueballs so these are not pionic dark matter models - we do not study such glueball dark 

\noindent matter models here. Region 9 is therefore intended as the transition regime where the pions are light but the glueballs are also playing a role in the IR dynamics.

Having obtained these results for the spectrum we will then return to discuss the phenomenology of the different types of IR theory in sec.~\ref{sec:phenomenology}. We stress here that in this work, we are only introducing this space of theories and our observations are initial and do not include relic density calculations which we hope to perform in the future led by this survey. 

The highlights of these analyses include the following. First, consider QCD-like models in which the low-energy spectrum contains only pions, while all other states lie above the $2\mpi$ threshold and can therefore spontaneously decay within the strong-sector. These models struggle to generate a large enough $\mpi/ \fpi$ value to be viable and need extra pion relic density generation modes such as through mediators with the Standard Model (SM). Our results suggest higher values of this ratio may be possible in very walking models. The largest space of models we find are those with an intermediate quark mass and they  have both the $\sigma$ and the $\rho$ lying below $2 \mpi$. These states may aid the pion depletion rate in the cosmological evolution as needed for valid theories. They are largely unexplored in the literature so far and we plan to return to them to determine relic abundances. Regions with just one of the $\rho$  or $\sigma$ light are possible in constrained parameter regions (obtaining a model with only a light $\rho$ is quite tuned in the holographic model).  The $\sigma$ always lies above the $\pi$ mass becoming degenerate with the $\pi$ in the extreme walking limit.

Finally in sec.~\ref{sec:dm_discussion}, to be complete in our survey, we briefly address other possible strongly coupled dark matter sectors. This includes the high mass scenario discussed above that leaves glueball dark matter \cite{daRocha:2017cxu,Carenza:2022pjd,Carenza:2023eua,Batz:2023zef,McKeen:2024trt,Soni:2016gzf}. If mesonic matter is unstable then dark baryons are possibly stable, see~\cite{Cline:2021itd,Kribs:2016cew} for associated reviews. For models truly in the chirally symmetric conformal window the lightest matter is an unparticle plasma \cite{Georgi:2007ek,Kikuchi:2007az}. Finally we  present some models that remove the pions from the spectrum leaving a dark matter $\sigma$ although in absence of additional symmetries it likely decays too fast due to dimension 5 operators to be a sensible dark matter candidate.

\section{The Holographic Model}
\label{sec:model}

The holographic model we use for the chiral symmetry breaking dynamics is presented in~\cite{Alho:2013dka}. It is similar in spirit to the  Anti-deSitter (AdS)/QCD models of~\cite{Erlich:2005qh,DaRold:2005mxj} but in addition includes the explicit running of the gauge theory and the chiral symmetry breaking is dynamically determined rather than input and fitted. A comparison of the model to SM QCD can be found in~\cite{Clemens:2017udk} - it is good at describing the broad features of the light mesonic spectrum but quantitatively can differ at the 20$\%$ level. Our goal is to use its qualitative power to elucidate what phases one can expect to find in such gauge theories as a function of $\nc,\nf$ and a common quark mass $m_Q$. It is also useful to see how the various low energy parameters depend on $\nc,\nf$. We will briefly present the action and equations of motion which will be used to calculate meson masses and decay constants.

The gravity action includes the bulk fields: the scalar $X$, which is dual to the operator $\bar{q} q$; its phase, which is dual to $\bar{q}\gamma_{5} q$; and the flavour gauge field $V^{M}$ (appearing in $F^{MN}$), which is dual to the vector current $\bar{q}\gamma^{\mu} q$. The action takes the form
\begin{eqnarray} \label{eq:general_action}
S_{boson} &=&  \int d^5 x ~ \varrho^3 \left( \frac{1}{r^2} (D^M X)^{\dagger} (D_M X)  
+ \frac{M_X^2}{\varrho^2} |X|^2  + \frac{1}{2 g_{5}^2} F_{MN}F^{MN} \right) .
\end{eqnarray}

The five-dimensional coupling only enters into quantities that explicitly compare different terms in the action - here for us this will just be the computation of $\fpi$ as we describe below. The fit to QCD in \cite{Clemens:2017udk} suggested $g_5^2=76$ and we will use that value here\footnote{We take $g_5$ to be scale, $\nf$, and $\nc$ independent by ansatz here.}. 

The model has a five-dimensional asymptotically AdS spacetime, the metric for which is
\begin{align}
ds_R^2 = r^2 dx^2_{(1,3)} + \frac{d r^2}{r^2} \, ,
\end{align} 
where $r^2 = \varrho^2 +|X|^2$ -- $\varrho$ is the holographic radial direction corresponding to the energy scale, and with the AdS radius set to one. The $X$ vacuum expectation value is included as a back-reaction in the metric.  

The dynamics of the particular gauge theory choices of $\nc, \nf$, with quark contributions to any running coupling, are included through $M_X^2$ in eqn.~\eqref{eq:general_action}. We use the perturbative result for the running of the anomalous dimension of the quark mass, $\gamma$ and expand the usual holographic relation between the mass of a scalar $M$ in AdS$_5$ and the operator dimension $\Delta$ -- $M^2=\Delta(\Delta-4)$ for a dimensionless scalar \cite{Witten:1998qj} -- at small $\gamma$ giving 
\begin{align} \label{dm}
 M_X^2 = - 2 \gamma.
\end{align}
Since the true running of $\gamma$ is not known non-perturbatively, we extend the perturbative results as a function of the renormalization group (RG) scale $\mu$ to the non-perturbative regime. We then directly set the field theory RG scale $\mu$ equal to the holographic RG scale $r= \sqrt{\varrho^2+|X|^2}$.
The model breaks chiral symmetry when $\gamma$ passes 1/2, as the Breitenlohmer-Freedman (BF) bound~\cite{Breitenlohner:1982jf} is then violated.

The two-loop result for the running coupling, $\alpha(\mu)$ is
\begin{align}
\mu \frac{d \alpha(\mu)}{d \mu} = - b_0  \alpha(\mu)^2 - b_1  \alpha(\mu)^3 \, ,
\end{align} 
with 
\begin{equation}   \label{running}
\begin{array}{ccl}
b_0 &=& \frac{1}{6 \pi} \left(11 C_A - 4  T_R \nf \right) \, ,\\ &&\\
b_1 &=& \frac{1}{24 \pi^2} \left(34 C^2_A - \left(20 C_A + 12 C_F \right) T_R\nf \right)  \, .
\end{array}
\end{equation}
where $T_R$ is half the Dynkin index, $C_F = (\nc^2-1)/2\nc$ the quadratic Casimir for the adjoint representation and $C_A = \nc$.

With our definition, the running coupling has IR poles at low $\nf$, IR fixed points at intermediate $\nf$ and describes the lower edge of the conformal window near $\nf = 4 \nc$ (it follows the choices in \cite{Appelquist:1996dq}). The true value of $\nf = N^c_f$ corresponding to the position of the edge of the conformal window is still being investigated and lattice, functional renormalization group studies suggest it is lower than this e.g., below  $N^c_f = 10$ for $\nc = 3$~\cite{Hasenfratz:2023wbr,Goertz:2024dnz}. Our eventual phase structure diagram fig.~\ref{fig:boxes_1}(right panel) though is shown in $\nf/N^c_f$ and we expect it to be broadly similar if the edge is lower. For $\nf > N^c_f$ the theory enters the conformal window and describes unparticles, which we briefly return to in sec. \ref{sec:dm_discussion}.

For the results to come, one must numerically  set $\alpha$ at some scale -- we set $\alpha=0.65$ at $\mu=1$.  We are careful though to rewrite all our results in units of the $\rho$ mass in the theory with $m_Q=0$ to remove this arbitrary choice. We will denote this $\rho$ mass as $M^0_\rho$ throughout this work.

Although our running coupling is computed at two loop, we use the one-loop anomalous mass dimension ($\gamma$), 
\begin{align} \label{grun}
\gamma(\mu) = \frac{3C_F}{2 \pi}~\alpha(\mu).
\end{align}
We use the one loop definition of $\gamma$ since it is already a guess non-perturbatively and no additional qualitative features are added beyond.

To find the vacuum of the theory,  we set all fields to zero except for $|X| = L(\varrho)$. For $M_X^2$ a constant, the equation of motion we obtain from eqn.~\eqref{eq:general_action} is
\begin{align}
\partial_{\varrho} (\varrho^3 \partial_{\varrho} L(\varrho)) - \varrho ~ M_X^2 L(\varrho) = 0 \label{eq: vacuum qcd} \, .
\end{align}
At large $\varrho$, in the UV, the asymptotic solution is $L(\varrho) = m_Q + c/\varrho^2$, with $c=\langle\Bar{q}q\rangle$, the fermion condensate of dimension three, and $m_Q$, the mass of dimension one. We numerically solve eqn.~\eqref{eq: vacuum qcd} with our input $M_X^2$ for the function $L(\varrho)$.

We use IR boundary conditions where the fermions go on mass-shell 
\begin{align} \label{vacIR}
L(\varrho)|_{\varrho=\varrho^{IR}} = \varrho^{IR} \, ,&& \partial_{\varrho} L(\varrho)|_{\varrho=\varrho^{IR}} = 0 \, .
\end{align}
The value of $\varrho^{IR}$ is determined in each theory. We numerically vary $\varrho^{IR}$ until the value of $L$ at the boundary is the desired fermion mass $m_Q$. We refer to the vacuum solutions as $L_{0}(\varrho)$.

The mesons of the theory are linearized fluctuations of this vacuum configuration that satisfy the appropriate boundary conditions, matching those of the vacuum in the IR and consisting of just fluctuations of operators in the UV. The resulting Sturm-Liouville problems fix the meson masses.

The vector-mesons are  fluctuations of the gauge field and satisfy the equation of motion (here $r^2= \varrho^2 + L_0(\varrho)^2$)
\begin{align} \label{eq: eqn of motion_vector}
\partial_{\varrho} (\varrho^3 \partial_{\varrho}V(\varrho)) + M^2_{\rho} \frac{\varrho^3}{r^{4}} V(\varrho) = 0.
\end{align}
To obtain a canonically normalized kinetic term for the vector meson we must impose (note this is $\nc,\nf$ independent) 
\begin{equation}  
\int d\varrho~ {\varrho^3 \over g_5^2 r^4} V(\varrho)^2 = 1. \label{Vnorm}
\end{equation}

The fluctuations of $L(\varrho)$ give rise to the scalar $\sigma$ meson \cite{Evans:2013vca}. The equation of motion for the fluctuation reads

\begin{equation} \label{eq: eqn of motion_scalar}
\partial_{\varrho} (\varrho^3 \partial_{\varrho} S(\varrho)) - \varrho (M_X^2) S(\varrho) - \varrho L_{0}(\varrho) S(\varrho) \frac{\partial M_X^2}{\partial L} |_{L_{0}} + M_\sigma^2 \frac{\varrho^3}{r^{4}} S(\varrho) = 0.
\end{equation}
The equation of motion for the phase of $X$ describes the pion field 
\begin{equation}
\partial_{\varrho} \left( \varrho^3 ~ L_0^2 ~ \partial_{\varrho} \pi \right) + \mpisq \frac{\varrho^3~L_0^2}{r^4} \pi   = 0  \, .\label{eq: 4}
\end{equation}

To compute decay constants, we must couple the meson to an external source and we include an axial vector field $A^M$ in analogy to $V^M$. In addition, since $X$ carries axial charge, there is an interaction term via the covariant derivative. Writing $A(\varrho)$ as the $\varrho$ dependent piece of  $A^M$,  it  has equation of motion (note this is where $g_5^2$ enters although the results for $\fpi$ are very insensitive to it's value). 
\begin{equation}  \label{eq: eqn of motion_axial}
\partial_{\varrho} (\varrho^3 \partial_{\varrho} A(\varrho)) - g_5^2 \frac{\varrho^3 L^2_{0}}{r^2} A(\varrho) - \frac{\varrho^3 q^2 }{r^{4}} A(\varrho) = 0\, . 
\end{equation}
The sources are described as fluctuations with a non-normalizable UV asymptotic form. We fix the coefficient of these solutions
by matching to the gauge theory in the UV.  In the UV we expect 
$L_0(\varrho) \sim 0$ and we can solve
the equations of motion for the scalar, $L= K_S(\varrho)$, vector $V^\mu= \epsilon^\mu K_V(\varrho) $, and  axial $A^\mu= \epsilon^\mu K_A(\varrho)$ fields. Each satisfies the same UV asymptotic equation 
\begin{equation}  \label{thing}
\partial_\varrho [ \varrho^3 \partial_\varrho K] - {q^2 \over \varrho} K= 0\,. 
\end{equation}
The solution is
\begin{equation} \label{Ks}
K_i = N_i \left( 1 + {q^2 \over 4 \varrho^2} \ln (q^2/ \varrho^2) \right),\quad (i=S,V,A),
\end{equation}
where $N_i$ are normalization constants that are not fixed by the linearized equation of motion.
Substituting these solutions back into the action gives the vector correlator $\Pi_{VV}$ and axial vector correlator $\Pi_{AA}$. Performing the usual matching to the UV gauge theory  requires us to set \cite{Erlich:2005qh,Alho:2013dka} 
\begin{equation}  \label{eq: match}
 N_V^2 = N_A^2 = {g_5^2 ~ d(R) ~ \nf(R) \over 24 \pi^2}, \hspace{0.5cm}  N_S^2 = {d(R) ~ \nf(R) \over 24 \pi^2 }. 
\end{equation}
where  $d(R)$ is the dimension of the representation ($\nc$ for the fundamental representation matter.)

The vector meson decay constant is then given by the overlap term between the meson and the external source
\begin{equation} F_V^2 = \int d \varrho {1 \over g_5^2} \partial_\varrho \left[- \varrho^3 \partial_\varrho V\right] K_V(q^2=0)\,.
\label{rhodecay}
\end{equation} 
Note the normalizations of the normalizable and non-normalizable fields combine to remove any $g_5^2$ dependence. $F_V^2$ scales as $\sqrt{\nc \nf}$. The process and scalings are similar for the scalar decay constant $F_\sigma$ for decays to an external scalar source (note this is not the coupling to an external spin two source considered as the dilaton decay constant \cite{Zwicky:2023krx}).

The pion decay constant can be extracted from the expectation that $\Pi_{AA} = f_\pi^2$ here in the QCD conventions i.e. $\fpi=93$MeV.  

\begin{equation} 
f_\pi^2 = \int d \varrho {1 \over g_5^2}  \partial_\varrho \left[  \varrho^3 \partial_\varrho K_A(q^2=0)\right] K_A(q^2=0)\,.
\end{equation} 
Explicit $g_5^2$ dependence again cancels against the normalization but $g_5^2$ enters through the solution for $K_A$. The scaling is $f_\pi^2 \sim \nc \nf$ now since two non-normalizable solutions enter.

We numerically solve the equations of the holographic model using {\tt NDSolve} in {\tt Mathematica} for the results below.

\section{Mass spectrum}
\label{sec:mass_spec}

We now present results and fits to the data from the holographic model. We begin with the SU(3) gauge theory where the results for the light meson spectrum and decay constants along with their fits are shown in fig.~\ref{fig:spectrum_nf_2} --~\ref{fig:spectrum_nf_11} and tab.~\ref{tab:mass_fits} --~\ref{tab:fit_decay_const}. Finally, we show the $\nf$ dependence of our fits in fig.~\ref{fig:spectra_nf_dependence}.

\subsection{$\nc=3$ results}

\begin{figure}[ht]
\centering
\includegraphics[width=.45\linewidth]{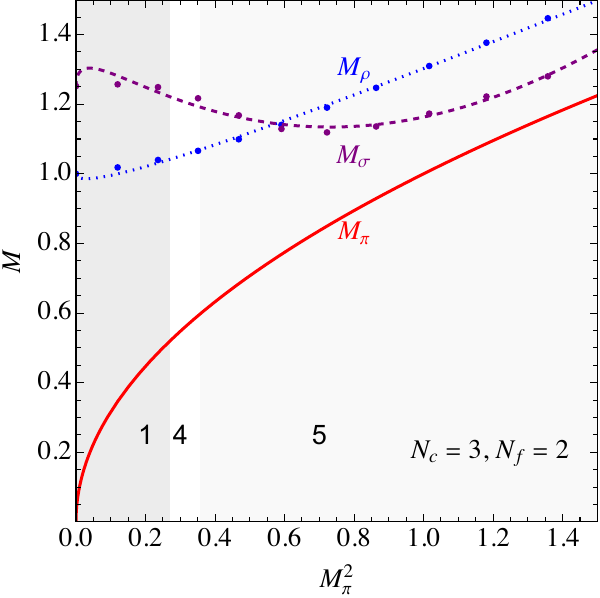}
\includegraphics[width=.45\linewidth]{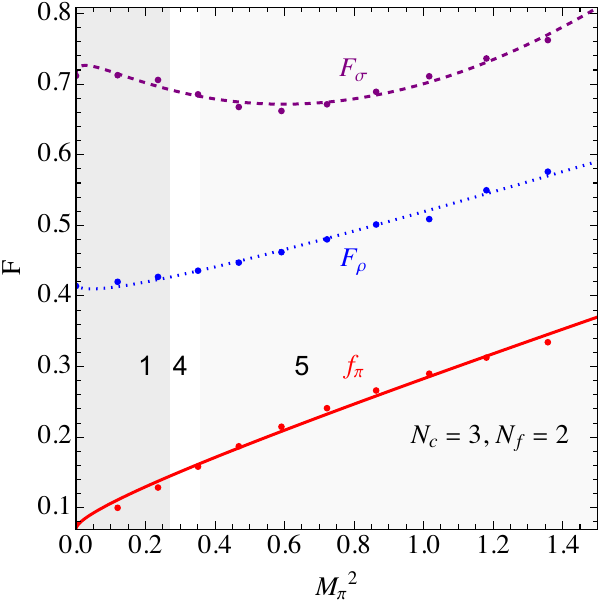}
\caption{Mass spectrum (left) and decay constants (right) in units of $M^0_\rho$ for $\nf = 2$. The gray shaded areas correspond to the different regions of parameter space presented in fig.~\ref{fig:boxes_1}.}
\label{fig:spectrum_nf_2}
\end{figure}

\begin{figure}[ht]
\centering
\includegraphics[width=.45\linewidth]{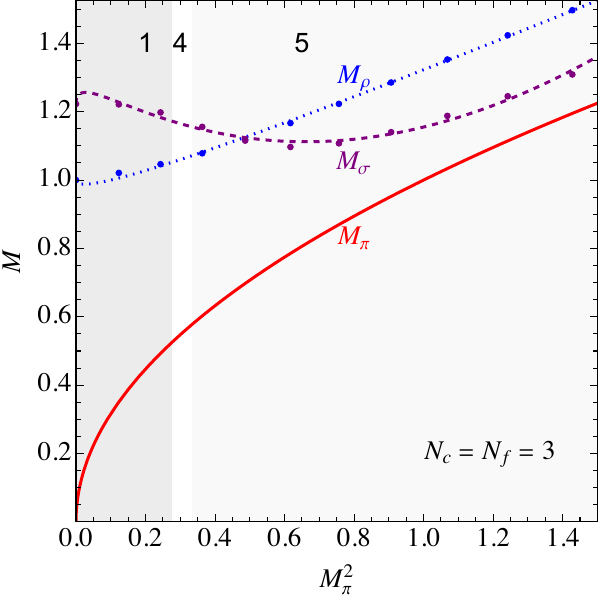}
\includegraphics[width=.45\linewidth]{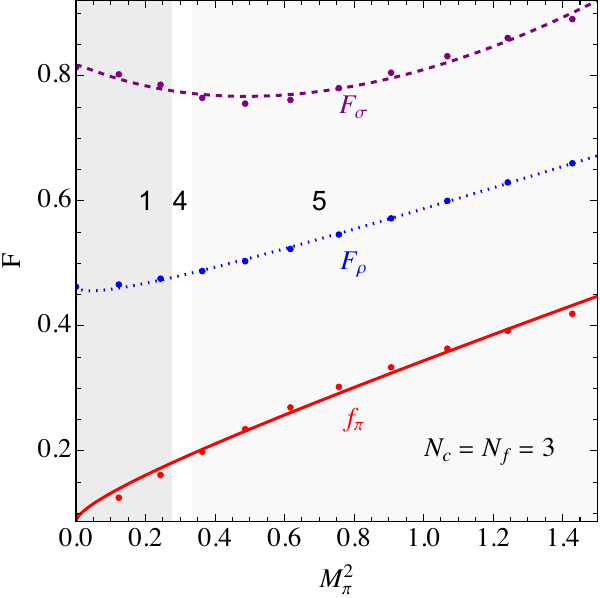}
\caption{Mass spectrum (left) and decay constants (right) in units of $M^0_\rho$ for $\nf = 3$. The gray shaded areas correspond to the different regions of parameter space presented in fig.~\ref{fig:boxes_1}. }
\label{fig:spectrum_nf_3}
\end{figure}

\begin{figure}[ht]
\centering
\includegraphics[width=.45\linewidth]{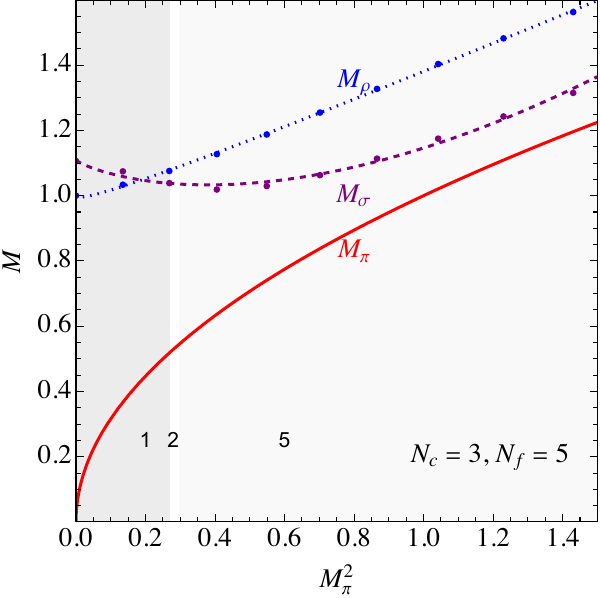}
\includegraphics[width=.45\linewidth]{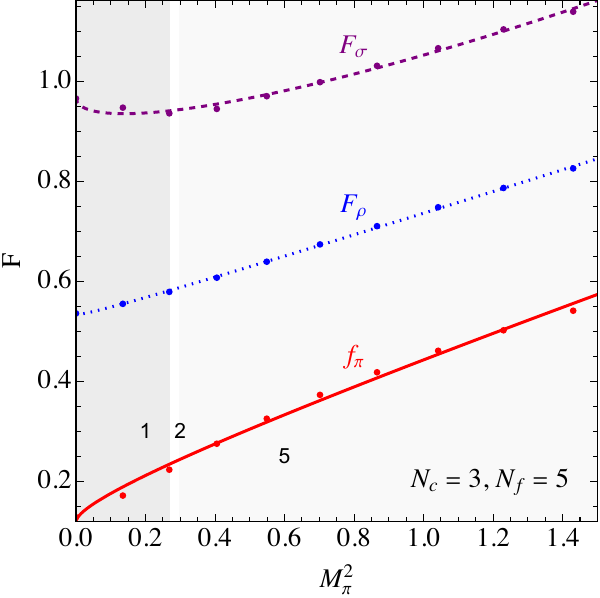}
\caption{Mass spectrum (left) and decay constants (right) in units of $M^0_\rho$ for $\nf = 5$. The gray shaded areas correspond to the different regions of parameter space presented in fig.~\ref{fig:boxes_1}.}
\label{fig:spectrum_nf_5}
\end{figure}

\begin{figure}[ht]
\centering
\includegraphics[width=.45\linewidth]{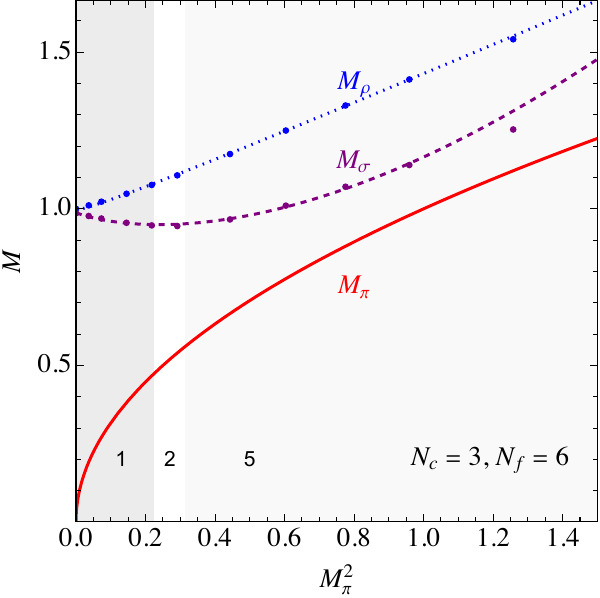}
\includegraphics[width=.45\linewidth]{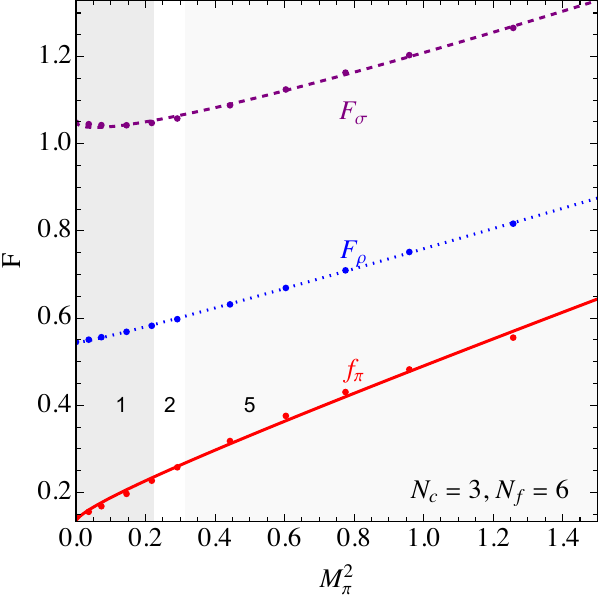}
\caption{Mass spectrum (left) and decay constants (right) in units of $M^0_\rho$ for $\nf = 6$. The gray shaded areas correspond to the different regions of parameter space presented in fig.~\ref{fig:boxes_1}.}
\label{fig:spectrum_nf_6}
\end{figure}
\begin{figure}[ht]
\centering
\includegraphics[width=.45\linewidth]{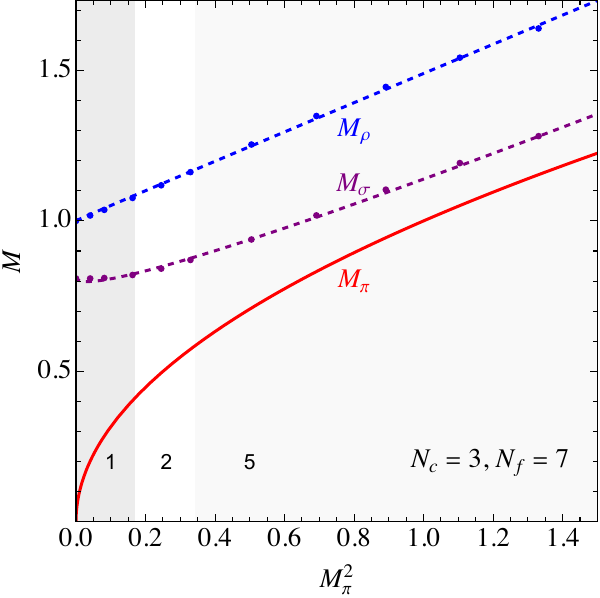}
\includegraphics[width=.45\linewidth]{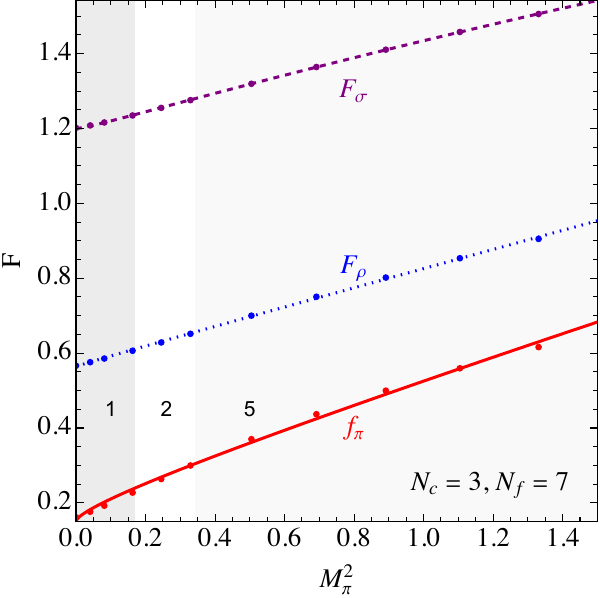}
\caption{Mass spectrum (left) and decay constants (right) in units of $M^0_\rho$ for $\nf = 7$. The gray shaded areas correspond to the different regions of parameter space presented in fig.~\ref{fig:boxes_1}.}
\label{fig:spectrum_nf_7}
\end{figure}

\begin{figure}[ht]
\centering
\includegraphics[width=.45\linewidth]{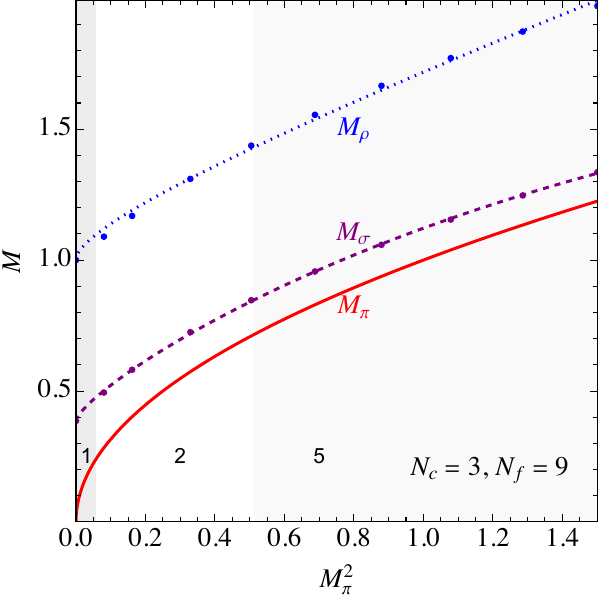}
\includegraphics[width=.45\linewidth]{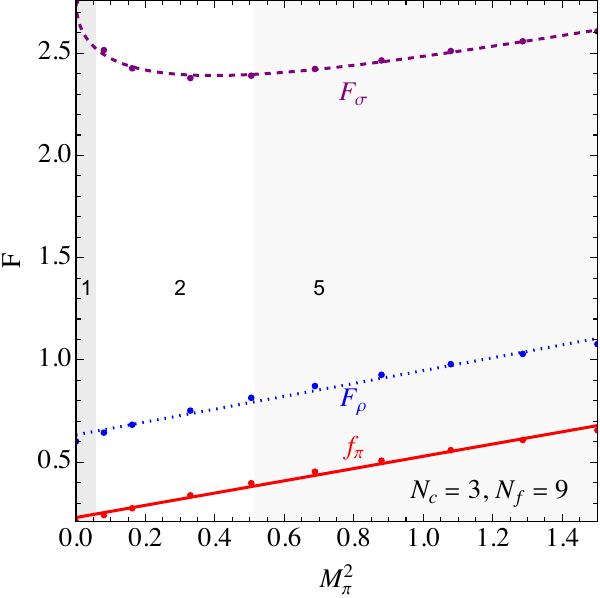}
\caption{Mass spectrum (left) and decay constants (right) in units of $M^0_\rho$ for $\nf = 9$. The gray shaded areas correspond to the different regions of parameter space presented in fig.~\ref{fig:boxes_1}.}
\label{fig:spectrum_nf_9}
\end{figure}
\begin{figure}[ht]
\centering
\includegraphics[width=.45\linewidth]{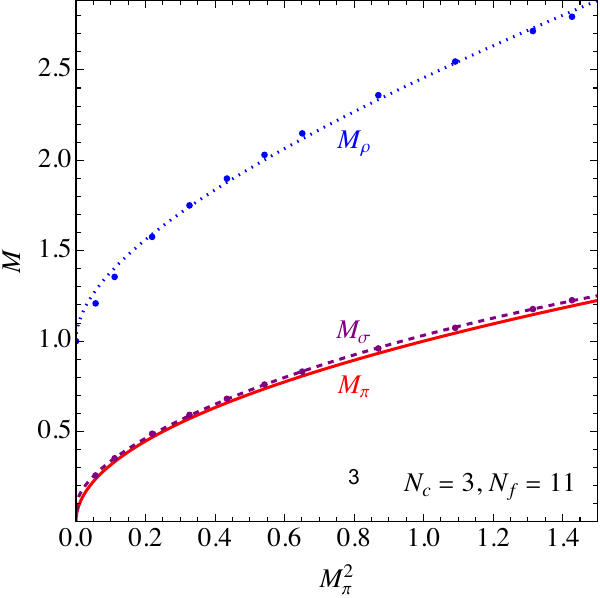}
\includegraphics[width=.45\linewidth]{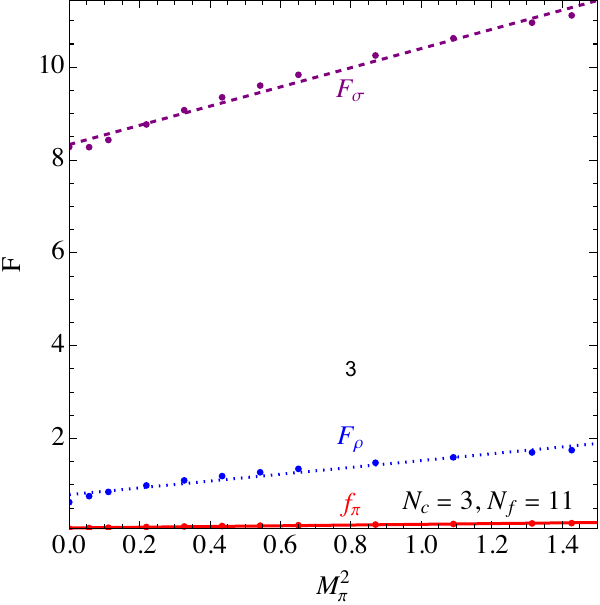}
\caption{Mass spectrum (left) and decay constants (right) in units of $M^0_\rho$ for $\nf = 11$. The gray shaded areas correspond to the different regions of parameter space presented in fig.~\ref{fig:boxes_1}.}
\label{fig:spectrum_nf_11}
\end{figure}

\begin{table}
\centering
\begin{tabular}{ |c|l|l| } 
 \hline
 $\nf$ & $M_\rho/M^0_\rho$ & $M_\sigma/M^0_\rho$ \\ 
 \hline
 2 & $1 - 0.161\mpi +0.463\mpisq$ & $1.25 + 0.604\mpi - 1.88M^2_\pi + 1.19\mpicube$ \\ 
 
 3 & $1 - 0.15\mpi +0.472\mpisq$ & $1.22 + 0.431\mpi - 1.56M^2_\pi + 1.06\mpicube$ \\ 
 
 5 & $1 - 0.097\mpi + 0.478\mpisq$ & $1.11 - 0.0313\mpi - 0.501M^2_\pi + 0.569\mpicube$ \\
 
 6 & $1 -0.063 \mpi + 0.495 \mpisq$ &  $0.988 - 0.002 \mpi - 0.488 \mpisq + 0.664 \mpicube$ \\ 
 
 7 & $1 + 0.00931 \mpi +  0.48 \mpisq$& $0.814 - 0.175 \mpi + 0.482 \mpisq + 0.0219 \mpicube$  \\
 
 9 & $1 + 0.309 \mpi + 0.409 \mpisq$ &  $0.384 + 0.201 \mpi + 0.846 \mpisq - 0.309 \mpicube$  \\ 
 
 11 & $1 + 1.09 \mpi + 0.363 \mpisq$ & $0.0736 + 0.715 \mpi + 0.431 \mpisq - 0.188 \mpicube$ \\
 \hline
\end{tabular}
\caption{Fits for $M_\rho, M_\sigma$ in units of $M^0_\rho$ as a function of $\nf$ for $\nc = 3$ derived within our holographic model.}
\label{tab:mass_fits}
\end{table}

\begin{table}
\centering
\begin{tabular}{ |c|l|l|l| } 
 \hline
 $\nf$ & $\fpi/M^0_\rho$ & $F_\rho/M^0_\rho$ & $F_\sigma/M^0_\rho$ \\ 
 \hline

2 & $0.07 + 0.070 \mpi + 0.143 \mpisq$ & $0.418 - 0.075 \mpi + 0.176  \mpisq$ & $0.71 + 0.186 \mpi - 0.69 \mpisq + 0.49 \mpicube$  \\ 
 
 3 & $0.086 + 0.094 \mpi + 0.164 \mpisq$ & $0.465 - 0.086 \mpi + 0.208  \mpisq$ & $0.82 + 0.04 \mpi- 0.42 \mpisq + 0.37 \mpicube$  \\ 
 
 5  & $0.119 + 0.114 \mpi + 0.211 \mpisq $ & $0.535 - 0.03 \mpi + 0.231  \mpisq$ & $0.97 - 0.17 \mpi + 0.19 \mpisq + 0.06 \mpicube$ \\
 
 6  & $0.133 + 0.093 \mpi + 0.265 \mpisq$ & $0.546 - 0.035 \mpi + 0.248  \mpisq$ & $1.05 - 0.13 \mpi + 0.27 \mpisq + 0.018 \mpicube$ \\ 
 
 7 & $0.149 + 0.11\mpi + 0.266 \mpisq$ & $0.564 + 0.012 \mpi + 0.25  \mpisq$ & $1.20 - 0.04 \mpi + 0.34 \mpisq - 0.06 \mpicube$ \\
 
 9 &  $0.229 + 0.3 \mpisq$ & $0.63 + 0.31  \mpisq$ & $2.76 - 1.27 \mpi + 1.24 \mpisq - 0.25 \mpicube$ \\ 
 
 11 & $0.072 + 0.078 \mpisq$ & $0.788 + 0.735  \mpisq$ &  $8.33 + 2.07 \mpisq$ \\
 \hline
\end{tabular}
\caption{Fits for decay constants $\fpi, F_\rho, F_\sigma$ in units of $M^0_\rho$ as a function of $\nf$ for $\nc = 3$ obtained using our holographic model.}
\label{tab:fit_decay_const}
\end{table}
The left hand plots in fig.~\ref{fig:spectrum_nf_2} -- \ref{fig:spectrum_nf_11} show the $\rho, \sigma$ and $\pi$ masses as a function of $\mpisq$ in the unit of $M^0_\rho$. At low quark mass $\mpisq \sim m_Q$ making  $\mpisq$ serve as an observable proxy for the quark mass. The red $\pi$ mass curve is, of course, trivial against $\mpisq$ but included for comparison to the other states. 

\begin{figure}[h!]
\centering
\includegraphics[width=.45\linewidth]
{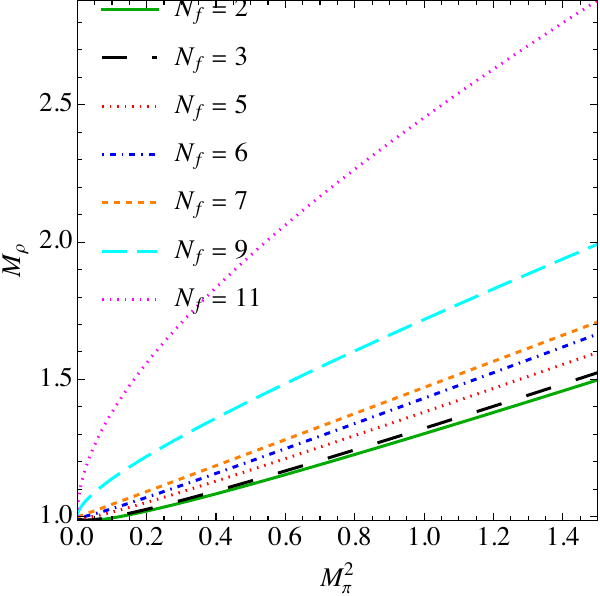}
\includegraphics[width=.45\linewidth]{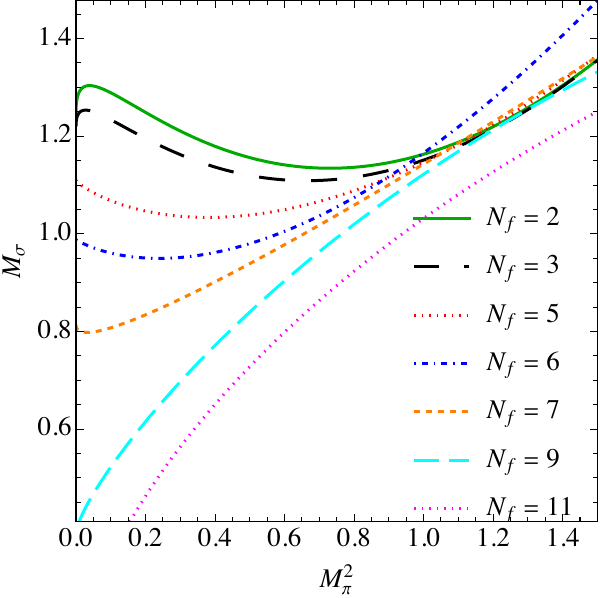}
\caption{$\nf$ dependence of the $\rho$ and $\sigma$ masses using the fits presented in table~\ref{tab:mass_fits}.}
\label{fig:spectra_nf_dependence}
\end{figure}

We normalize the mass scales of the spectrum by setting the $\rho$ mass at $\mpi = m_Q = 0$ to be unity in each case. Thus $M_\rho(m_Q=0) = M^0_\rho = 1$ is the measure of the strong coupling scale when one compares theories. The $\rho$ mass is well fitted as a function $a + b \mpi + c \mpisq$ at all $\nf$ and are shown in tab.~\ref{tab:mass_fits}. As shown in fig.~\ref{fig:spectra_nf_dependence} (left panel), the fits we obtain are a function of $\nf$, although we do not attempt to fit this dependence. The $\nf$ dependence of the $\rho$ mass becomes stronger as one approaches the lower edge of the conformal window at $\nf \sim 11$ (formally the edge lies at 11.9). The $\mpi$ term in the fits makes the largest difference near the edge of the conformal window. These results clearly demonstrate the importance of establishing $\nf$ dependence of heavier resonance properties by means of lattice simulations.

 We have concentrated on the regime $M^2_\pi < 1.5 (M^0_\rho)^2$ -- in the holographic model above this scale there is a transition to a linear regime where all $M,F \propto \mpi  \sim m_Q$ but in all but the most walking theories the gauge theory becomes perturbative here and  holography isn't the correct description. Also in these high $m_Q$ cases the glue dynamics separates from the quark dynamics below the quark mass scale. These theories will include glueballs that are lighter than the mesons so the models are no longer pion based dark matter models. We have not attempted to describe the glueball sector here but we note that at intermediate $m_Q$ there will be theories with glueballs as well as light $\rho, \sigma, \pi$. Without being quantitative about the intermediate $m_Q$ region, in fig.~\ref{fig:boxes_1} (left panel) we show this as region 9 which corresponds to intermediate quarks where glueballs will be a relevant light degrees of freedom. This region lies at higher $M^2_\pi$ in fig.~\ref{fig:boxes_1} (right panel).

The $\sigma$ meson mass is much more $\nf$ dependent than the $\rho$ (see fig.~\ref{fig:spectra_nf_dependence}, right panel), at least in this holographic model. The key term is the third term in eqn.~\eqref{eq: eqn of motion_scalar} which depends on the rate of change of the running mass $M_X^2$. Without this term the equation becomes degenerate with the pion eqn.~\eqref{eq: 4} (using eqn.~\eqref{eq: vacuum qcd}). Thus at the edge of the conformal window where the IR running of the coupling is negligible the $\sigma$ becomes degenerate with the $\pi$. On the other hand in the low $\nf$ theories the running is fast and the resulting $\sigma$ is much heavier (heavier than the $\rho$ for $\nf<5$).

We have used the fits to determine where the IR theory lies in the space of theories described in fig.~\ref{fig:boxes_1} (left panel) and this then determines the regions shown in fig.~\ref{fig:boxes_1} (right panel). We will discuss fig.~\ref{fig:boxes_1} (right panel) in detail below.

Finally, in fig.~\ref{fig:spectrum_nf_2}  -- \ref{fig:spectrum_nf_11} (right panels) we also show the corresponding decay constants for the $\rho, \sigma, \pi$. They all grow with $\mpi$. The main $\nf$ scaling is that discussed in section~\ref{sec:model} -- $f^2_\pi \sim \nf$ and $F_{\rho, \sigma}^2 \sim \sqrt{\nf}$. In addition there is $\nf$ dependence through the holographic wave functions of the particles involved. Here the biggest effect is in $F_\sigma$ which grows sharply with $\nf$ as one approaches the edge of the conformal window.

\subsection{$\nc$ Dependence}

At large $\nc$, i.e.\ the Veneziano limit~\cite{Veneziano:1968yb}, the $\beta$ function ansatz we use to input the dynamics of the theory is simply a function of $\nf/\nc$ at leading order. As is frequently argued $\nc=3$ lies close to large $\nc$ so it is interesting to test how good an approximation this scaling is. 

\begin{figure}[h!]
\includegraphics[width=.45\linewidth]{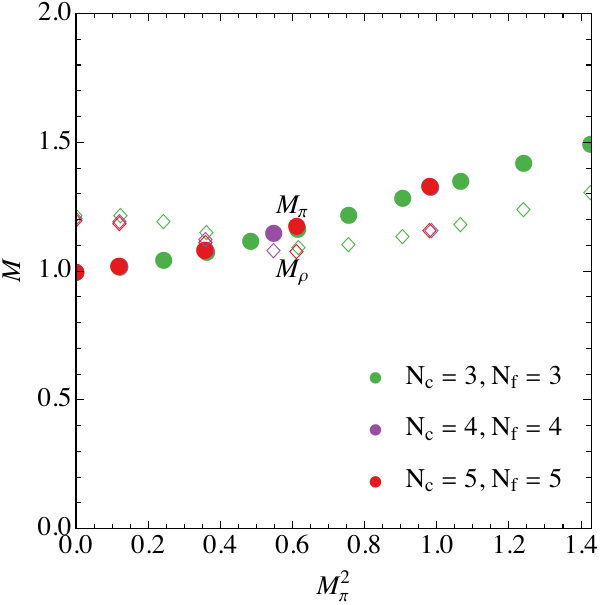}
\includegraphics[width=.45\linewidth]{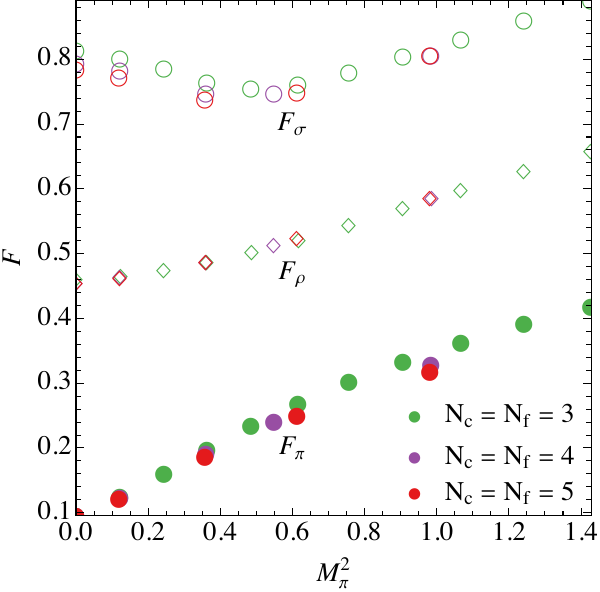}
\caption{Comparison of the theories   with $\nf=3$ (green), SU(4) with $\nf=4$ (Purple) and SU(5) with $\nf=5$ (Red) Left: data for the $\sigma$ and $\rho$ meson masses  Right: the decay constants $\fpi, F_\rho, F_\sigma$ corrected by the rude $\nf, \nc$ scaling discussed in the text.}
\label{fig:nc_scaling}
\end{figure}

 A natural set of theories to compare are SU(3) with $\nf=3$, SU(4) with $\nf=4$ and SU(5) with $\nf=5$. 
We show a selection of computations at various quark mass of the meson and decay constants in fig.~\ref{fig:nc_scaling}. To compare the decay constants we remove the explicit $\sqrt{\nc \nf}$ factors in $\fpi$ and $F_\rho^2, F_\sigma^2$. Certainly within any errors the holographic models contain, this scaling is clearly present. We have also cross checked SU(3) with $\nf=9$, SU(4) with $\nf=12$ and SU(5) with $\nf=15$ where again within 10\% the degeneracy is accurate. We will therefore not present further results 

\begin{figure}[h!]
\includegraphics[width=.475\linewidth]{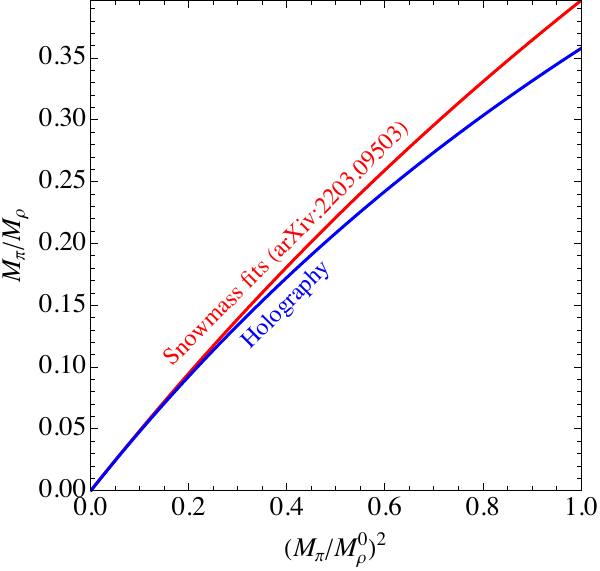}
\includegraphics[width=.45\linewidth]{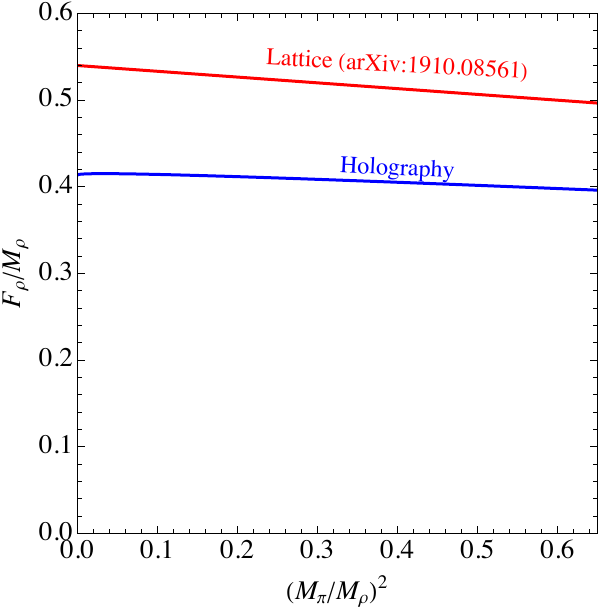}
\caption{ Left: Comparison of the holographic predictions (blue) for QCD compared to the lattice results in~\cite{Albouy:2022cin} (red) for $\mpi/M_\rho$ vs $\mpisq/M_\rho^{0~2}$ in units of the $\rho$ mass at $m_Q=0$. Right: a comparison of the holographic  results (blue) for $F_\rho/M_\rho$ vs $\mpisq$ in units of the $\rho$ mass (note not at $m_Q=0)$ and the lattice results from ~\cite{DeGrand:2019vbx} (red). }
\label{fig:lattice_comparisons}
\end{figure}

\noindent at higher $\nc$ since they can be extrapolated from $\nc=3$ and the corresponding $\nf$ variations. Similarly fig.~\ref{fig:boxes_1} (right panel) although valid for $\nc=3$  is in fact a good description of all $\nc$ theories.

\section{Comparison to Lattice Data}
It is worth making some comparisons of the holographic predictions to known lattice calculations. Several lattice investigations for SU($\nc$) gauge group are available. Most notably, ref.~\cite{Bali:2013kia,Bali:2008an,DelDebbio:2007wk} provide a comprehensive study of meson properties for large-$\nc$ theories in the quenched approximation. We choose here data from~\cite{DeGrand:2019vbx} for our comparisons as it is presented in the most useful fashion for our purposes.

The mass spectrum is the most important component of our analysis. In fig.~\ref{fig:lattice_comparisons} we show a comparison to  QCD functional methods analysis from~\cite{Albouy:2022cin,Fischer:2006ub} (here we take $\nc = \nf = 3$ in the holographic model to represent the light states in the 2+1 quark theory). The quantity $\mpi/M_\rho$ is plotted as a function of $\mpisq$ in units of $M^0_\rho$. At low $\mpisq$ this is a plot against the quark mass. The comparison is good. There is a modest deviation at larger $\mpisq$, which likely stems from the fact that in holographic models the conformal, weakly coupled UV of QCD is replaced by a strongly coupled conformal UV theory. Introducing a mass scale in the UV adds additional strong interactions so heavy quarks are poorly described. The holographic description of the $\rho$ seems very good up to $(\mpi/M_\rho^0)^2=0.5$ and even above there is decent agreement to draw qualitative conclusions. The transitions between regions in fig.~\ref{fig:boxes_1} (right panel) lie below this value. 

The dimensionful $F_\rho$ rises linearly with $\mpisq$. In the lattice review~\cite{DeGrand:2019vbx} they call what we call $F_\rho^2$, $M_V^2 F_V$. We fit their data and use our conventions. In fig.~\ref{fig:lattice_comparisons} we plot the dimensionless $F_\rho/M_\rho$ in the holographic model against the lattice data taken from~\cite{DeGrand:2019vbx}. The holographic model prediction is a little lower but the weak mass dependence is reasonably consistent. 

The holographic model's prediction for $\fpi$ is more troubling. We can see in the $\nf = \nc = 3$ case in fig.~\ref{fig:spectrum_nf_3} -- ~\ref{fig:spectrum_nf_5} that $\fpi$ approximately quadruples across the $\mpisq$ range. If one looks at the equivalent plot in~\cite{DeGrand:2019vbx} $\fpi$ does not even double across this range. We believe this is again a result of the UV artifice of the holographic model. As the quark mass rises above the strong coupling scale $\Lambda$ in QCD $\fpi$ is expected to scale as $\sqrt{m_Q  \Lambda}$ reflecting the role of the strong coupling scale still in the bound state dynamics. On the other hand in holography there is strong coupling at the scale $m_Q$ and one sees $\fpi \sim m_Q$. The physics in the holographic model is wrong and this shows up at lower quark mass with $\fpi$ rising more sharply to attain its larger UV quark mass behaviour. We nevertheless present the $\fpi$ predictions because they include interesting $\nf$ dependence that we believe is qualitatively correct and has lessons for the QCD-like theories. 

\begin{figure}
    \centering
    \includegraphics[width=.45\linewidth]{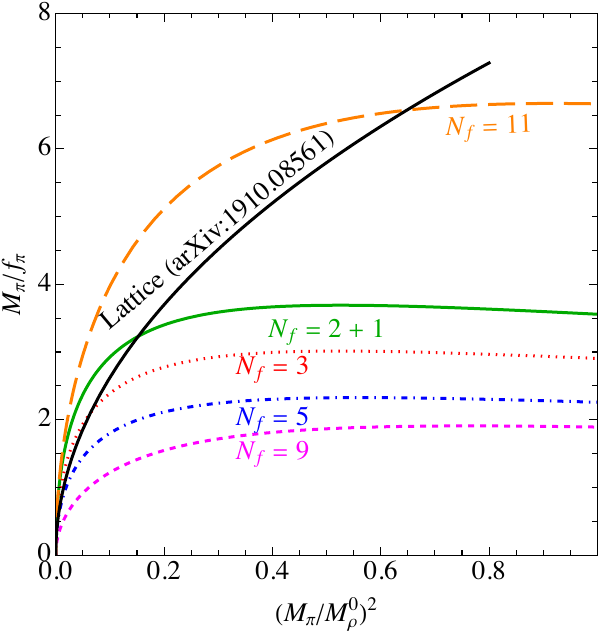} 
    \caption{The ratio $\mpi/\fpi$ in the holographic model as a function of $\mpi$ for $\nc = 3$, and various $\nf$ together with the lattice data (black line) from~\cite{DeGrand:2019vbx}. }
    \label{fig:lattice_comparison_mpifpi}
\end{figure}

Let us immediately turn to the quantity $\mpi/\fpi$ which plays an important role in SIMP models. In fig.~\ref{fig:lattice_comparison_mpifpi} we plot this quantity against $\mpisq/M_\rho^{2}$ - note here that it is $M_\rho$ in the denominator, not $M_\rho^0$ - the ultra heavy quark limit is at $\mpisq/M_\rho^2=1$. We plot the lattice results fitted from~\cite{DeGrand:2019vbx} and our holographic results for a number of choices of $\nf$. 

Here one should immediately compare the $\nf=2+1$ holographic theory (we use the $\nc=3$ running but $\nf=2$ in the decay constant calculation) and the lattice results. Until $\mpi/M_\rho \sim 0.5$  one could live with the holographic model but above this mass value the spurious UV behaviour of $\fpi$ kicks in and takes the holographic model away from the true lattice data. 

The lattice data shows that for $\nc = 3, \nf = 2+1$ reaching a value of $\mpi/\fpi$ greater than 5-6 seems very hard~\cite{DeGrand:2019vbx}. The original SIMP phenomenology wanted this ratio to be 12. The higher $\nf$ holography curves are lower reflecting the $\sqrt{\nc \nf}$ factor one expects in $\fpi$. Raising $\nc$ or $\nf$ will only make things worse ($\mpi/\fpi$ lower). This observation motivates our search for more complex models with additional dark matter relic density mechanisms present. 

The only outlier is the $\nf = 11$ curve which does reach higher values of the ratio. This theory is very walking and lives in Region 3 which we discuss below. The effect of walking (for which the idea was invented in \cite{Holdom:1981rm}) is to raise the quark condensate relative to $\fpi$. This rise in the holographic model infects all other masses and decay constants and hence $\mpi / \fpi$ can be larger. In the holographic model the ratio rises indefinitely as one tunes to $\nf^c$. This conclusion is likely to map across from the holographic model with poor UV behaviour to true walking SIMP models which therefore might have an important role to play in the SIMP paradigm.

\section{Phenomenology}
\label{sec:phenomenology}

A strongly coupled dark matter sector, such as the $SU(\nc)$ gauge theories we have considered here, in principle feature an infinite tower of states. The expectation is that for most of the parameter space the IR is well described by an effective theory of a few light meson states that all other states decay to.  Our main goal here has been to determine the likely states in those IR theories for SU($\nc$) dynamics as a function of $\nc,\nf$ and a common quark mass. As we have seen the spectrum is largely $\nc$ independent at fixed $\nf/\nc$ although the decay constants grow with $\nc$. We have identified 7 viable low energy theories shown in fig.~\ref{fig:boxes_1}. We have displayed the phase space in the $\nf$ vs $m_Q$ plane where each occurs in fig.~\ref{fig:boxes_1} (right panel). Here one can interpret the area in the plane as indicative of how tuned the parameter space for each region is. Of course the reliability of those conclusions depends on one's trust of the holographic model. The $\rho$ mass seems perfectly reasonably modelled (see fig.~\ref{fig:lattice_comparisons}). The $\sigma$ meson seems more model dependent. The holographic model places it heavier than the $\rho$ at low $\nf$ and low $m_Q$ (the identification of the $\sigma$ meson among the $f_0$ states in QCD is still uncertain)
before becoming lighter for larger $\nf$ as it moves towards a dilaton behaviour in the walking regime. Whilst this behaviour seems reasonable it is not a first principles result. At worst we view the holographic model as a guide to the types of phases that may exist.

It is worth noting the trajectory of the theories through the regions as $\nf$ and $m_Q$ change for a fixed $\nc$: 
\begin{enumerate}
    \item {\it Massless theories ($m_Q = 0$)}: In the chiral limit, as $\nf$ increases one moves from theories containing only light $\pi$s (region 1) $\rightarrow$ light $\pi, \sigma$ (region 2) $\rightarrow$ walking dilaton $\sigma$ (region 3) before entering the regime of unparticles \cite{Georgi:2007ek} in the conformal window.
    \item {\it Increasing mass ($m_Q > 0$), small $\nf$}: Low $\nf$ theories move from only light $\pi$s (region 1)  $\rightarrow$  light $\pi, \rho$  (region 4) $\rightarrow$ light $\pi, \rho, \sigma$  (region 5)  $\rightarrow$ light glueballs (region 9), although region 4 is rather a tuned part of the parameter space. 
    \item {\it Increasing mass ($m_Q > 0$), large $\nf$}:  For higher $\nf$ one sees light $\pi$s (region 1) $\rightarrow$ light $\pi, \sigma$ (region 2)  $\rightarrow$ light $\pi, \rho, \sigma$ $\rightarrow$ light glueballs (region 9) or walking dilaton $\sigma$ (region 3) $ \rightarrow$ light $\rho, \sigma$  (region 6) $\rightarrow$ light glueballs (region 9) at the edge of the conformal window.
    \item {\it No Regions 7,8:} there is no mechanism we know to make the $\rho$ anomalously light relative to the $\sigma$ and $\pi$ (whereas for those we can use chiral symmetry breaking or conformal symmetry breaking to make them light). 
\end{enumerate}

We will now turn to discuss the phenomenology, as currently known, for these seven regions featuring different mass hierarchies although we do not aim to rigorously analyse this phenomenology here. We also stress that we do not consider dark matter -- SM thermalization aspects, for which an external mediator between the strong-sector and the SM is often introduced. Our aim here is to analyse the strong sector in isolation and consider potential impact of higher dimensional effective couplings between the SM and strong sector which must arise at least at the Planck scale.

\subsection{Region 1: only light $\pi$}
Region 1 describes QCD-like theories with a strongly running coupling and light quark mass relative to the strong coupling scale. The spectrum including the $\sigma$ and $\rho$ lie above twice the pion mass at the strong coupling scale and so will decay quickly to pions. The pions are light because of their pseudo-Goldstone nature associated with the spontaneous breaking of chiral symmetry. Thus the only relevant IR dynamical degree of freedom in the system are pions which are described by a chiral Lagrangian for dark matter calculations. One expects effective dark pion - SM couplings. The lowest dimension examples are the UV dimension-7 operator $\bar{Q} \gamma_5 Q \bar{q} \gamma_5 q h$  ($h$ is the Higgs, $Q$ are dark quarks and $q$ SM quarks) and the UV dimension-8 operator $h^2 (\bar{Q} \gamma_5 Q)^2$.  These are highly suppressed by the ultraviolet scale and pion annihilations to the SM will thus be negligible allowing strong-sector only interactions to dominate.

The relic abundance in the pion model has been computed in~\cite{Hochberg:2014kqa}. Famously, the presence of a Wess-Zumino-Witten term of the form 
\begin{equation}
    {\cal L_{WZ}} \propto {1 \over  \fpi^5} \epsilon^{\mu \nu \rho \sigma} {\rm Tr} [\pi \partial_\mu \pi \partial_\nu \pi \partial_\rho \pi \partial_\sigma \pi]
\end{equation}
generates $3\to 2$ processes that deplete the pions and lead to lower dark matter masses than in other models. The authors traded the strong coupling scale and $m_Q$ for $\mpi$ and the ratio $\mpi/\fpi$. The pion mass is predicted to lie in the ten-few hundred MeV for $\mpi/\fpi$ up to 12 (Here again we stress we use the convention where $\fpi=93$MeV)\footnote{Note that we use a definition of $\fpi$ that differs by $2$ relative to~\cite{Hochberg:2014kqa} -- this figure is with our definition.}. 

Equally though the model contains four point self interactions of the pions (coming from the Tr$[M^\dagger U]$ type terms in the chiral Lagrangian ($ U=\exp(\pi/\fpi)$) in low energy limit where one neglects derivative interactions). The cross-section behaves as
\begin{equation}
    \sigma_{\rm scatter} \propto {\left(\mpi \over \fpi\right)^4} {1 \over \mpisq}
\end{equation}
so that at fixed $\mpi/\fpi$ one needs a sufficiently large pion mass for the self interaction to explain but not violate the bound from the bullet cluster observations~\cite{Wittman:2017gxn,Randall:2008ppe,Robertson:2016xjh}. 

The conclusion was that the dark pion mass should lie in the few 100s MeV with large $\mpi/\fpi \sim 12+$. This value lies close to the naive perturbativity bound of the chiral Lagrangian and implies a rather large quark mass whilst maintaining the chiral Lagrangian structure~\cite{Hansen:2015yaa,Braat:2023fhn,Pomper:2024otb}. We have seen in fig.~\ref{fig:lattice_comparison_mpifpi} above that neither lattice data nor our holographic models suggest it is easy to achieve such a large value. This suggests the other regions we discuss below may be of more phenomenological interest. Apart from the conventional $3\to 2$ mechanism discussed above, dark pion relic density can be generated through multiple mechanisms: via bound states of two pions, as demonstrated in~\cite{Chu:2024rrv}, via interactions among stable and transient pions in the presence of t‑channel mediators and residual flavor symmetries, as in~\cite{Carmona:2024tkg} or via the misalignment mechanism~\cite{Alexander:2023wgk} if the dark pions are very light. Finally, dark pion - SM couplings may also be generated via non-trivial topological structures~\cite{Davighi:2024zip, Davighi:2025awm} and may help generate the right relic density.

\subsection{Region 2: light $\pi, \sigma$ }

Region 2  is a substantial part of the parameter space in fig.~\ref{fig:boxes_1} (right panel) at larger $\nf$ values and intermediate quark masses. In these models the running of the coupling at the scale of the chiral condensate is weaker than in QCD (either because of the larger $\nf$ or the quark mass scale). The holographic model then predicts a light $\sigma$ meson below twice the pion mass. 

The $\sigma$s therefore can not decay directly to two pions and naively will form part of the final dark matter mix. However one can expect couplings between the quantum number free $\sigma$ and the SM from the UV completion of the theory. For example, a $c_{\sigma hh} \sigma |h|^2$ Lagrangian term where $c_{\sigma hh}$ is the coupling to the SM Higgs. The dark matter quark anti-quark operator that forms the $\sigma$ is dimension 3. Thus $c_{\sigma hh}$ will be given by the ratio of the dark sectors strong coupling scale squared to the  scale at which the dark sector and SM interact. 
This coupling between the $\sigma$ and $h$ will induce a mixing between these particles leading to the decay of $\sigma$ into two fermions. The partial widths are of the form
\begin{align}
\Gamma(\sigma \to f \bar{f}) \simeq \frac{1}{8 \pi m^2_\sigma} Y^2_f \left(\frac{c_{\sigma h h}}{\lambda v}\right)^2 (m^2_\sigma - 4 m_f^2)^{3/2}
\end{align}
with  $Y_f$ the usual Yukawa couplings,$\lambda$ the usual Higgs quartic coupling and $v$ the Higgs vacuum expectation value.
Due to the strong hierarchy in the Yukawa sector, the decay into the heaviest fermion kinematically allowed will determine the order of its life-time.
We assume here $m_\sigma \ll m_h$. Assuming the UV completion is at the Planck scale the width is approximately $Y_f^2 m_\sigma^5/v^2 \Lambda_{Pl}^2 \sim Y_f^2(1 {\rm GeV})^5/(246 {\rm GeV})^2/(10^{19} {\rm GeV})^2 \sim Y_f^2\, 10^{-42} \rm{GeV}$. Note, the age of the Universe is $ \sim 10^{41}\,\rm{GeV}^{-1}$. For a light dark $\sigma$ of order 1 GeV the decay can only proceed to lighter fermions than the top and the Yukawa suppression (e.g. for decays to the strange quark $Y_f^2 \sim 10^{-7}$) will make the $\sigma$ effectively stable and hence a dark matter candidate. Some avenues for dilaton dark matter have been explored in~\cite{Hong:2019nwd,Hong:2022gzo,Cho:2007cy,Cho:1998aa}. 

In any model where the UV completion provides thermalization to the SM sector the UV scale is likely much lower than the Planck mass and the $\sigma$ is unlikely to be stable and in fact can be expected to decay quickly. In such cases the $\sigma$ can still play a significant role in the evolution of the $\pi$. Strong interactions will allow the process ($2 \pi \rightarrow 2 \sigma$) during the regime where the kinetic energy of the $\pi$ is sufficiently large. The $ \pi \pi \sigma$ coupling together with $\sigma$ decaying to the SM effectively allow $\pi \pi \to f \bar{f}$ with a rate that depends on the UV completion scale.  Overall this may provide a depletion mechanism for the $\pi$. This has been analysed in~\cite{Appelquist:2024koa}. It will further be interesting to understand the $\nf$ dependence of this annihilation channel for which holography may provide qualitative guidance, but lattice computations are necessary for more quantitative estimates. The $\sigma$ meson may also alter the pion-interaction strength, therefore impacting the phenomenologically viable regions~\cite{Kondo:2022lgg}.

\subsection{Region 3: walking dilaton sigma }

Region 3 is the extreme walking limit of Region 2 where the running is so slow that the $\sigma$ meson becomes essentially degenerate with the pions. The holographic model suggests this occurs when $\nf$ is within approximately 10\% of $\nf^c$. This is when the new decay channel of Region 2 ($2 \pi \rightarrow 2 \sigma$) becomes maximally strong.

Furthermore as we saw in fig.~\ref{fig:lattice_comparison_mpifpi} it is possible to realise much large $\mpi / \fpi$ values in this region. The effect of walking (for which the idea was invented in~\cite{Holdom:1981rm}) is to raise the quark condensate relative to $\fpi$. This rise in the holographic model infects all other masses and decay constants and hence $\mpi / \fpi$ can be larger. 

Together, these effects should facilitate obtaining the observed relic density compared to the pure pion model, as long as the $\sigma$ decays sufficiently rapidly.

\subsection{Region 4: light $\pi, \rho$}
Phenomenologically, region 4 has been assumed to be a dominant region in parameter space of strong-sector theory. Our results in fig.~\ref{fig:boxes_1} show that region 4 is in fact only a small part of the parameter space. At small $\nf$ the $\sigma$ is heavy due to the strong running. As the pion mass rises it reaches half the $\rho$ mass first, however $\sigma$ quickly becomes light enough to be phenomenologically significant. 

Within region 4, the $\rho$ can be created for example by $3 \pi \rightarrow \pi \rho$ processes, but cannot quickly decay back to pions. If $\rho$ decays to the SM are rapid enough, this channel can be used to deplete the pion abundance efficiently as it destroys two pions per annihilation. This mechanism also allows one to lower the required $\mpi/\fpi$ to reconcile the relic density and self-scattering cross-section. This scenario has been studied in~\cite{Bernreuther:2023kcg}. 
In addition to $3 \pi \rightarrow \pi \rho$ processes $\pi \pi \to \rho \rho$ processes are also a possible relic density mechanism, given that $\rho$ decays are rapid. A possible model is presented in~\cite{Bernreuther:2019pfb}. Additionally the $\rho$ may also take part in the $3\to 2$ number changing processes modifying the phenomenologically viable region~\cite{Choi:2018iit}.

\subsection{Region 5: light $\rho, \sigma$}
Region 5 is the largest region in fig.~\ref{fig:boxes_1} (right panel). It occurs at all $\nf$ and when the quark mass is sufficiently large that the $\pi$ mass is more than about $0.4M^0_{\rho}$. Here the $\pi$ is essentially heavy enough to bring the $\rho$ and $\sigma$ below $2 \mpi$. Both the $\rho$ and $\sigma$ can now provide additional annihilation channels to reduce the pion abundance. This scenario is likely to be even more successful at making a working model with low $\mpi/\fpi$. Note that here and in the next two regions other higher mass bound states such as the axial vector mesons may begin to become stable against decay to lighter states however that is beyond the scope of this work. 

We have not found discussion of this case in the literature. Several competing processes such as $3\pi \to \pi\sigma,3\pi \to \pi\rho$, $3\pi \to \rho\sigma$, $2\pi \to 2\rho$ and $2\pi \to 2\sigma$ may be present in this region. Depending on the model construction, this offers a rich signature space not only for relic density generation mechanisms but also for experimental analyses~\footnote{For a review of collider searches for strong-sector theories see~\cite{Albouy:2022cin}.}. We do not perform the relic density calculations or signature space analysis here but one of our key findings is that this scenario is a very likely one. It would be very interesting to study it in more detail.

We also note that low $\nc$, $\nf = 1$ theories are also essentially region 5 theories. for these theories, the strong effect of the axial anomaly makes the pNGB heavy (as the $\eta^\prime$ is in QCD). It is therefore likely that the $\eta^\prime,\rho$ and $\sigma$ will all be close in mass~\cite{Farchioni:2007dw}.

\subsection{Region 6: dilaton $\sigma$, light $\rho$}
Region 6 is simply a walking variant of Region 5 where the $\sigma$ is degenerate with the $\pi$ and the $\rho$ is light. It doesn't appear in fig.~\ref{fig:boxes_1} (right panel) because it lies on the right hand end of the $\nf=11$ (at $\nc=3$) green line above $\mpi=2M^0_\rho$. 

\subsection{Region 9: light glueballs}
Region 9 occurs in all theories when the quark mass rises to of order the strong coupling scale. The IR theory will consist of $\pi, \rho, \sigma$ and scalar glueballs. The scalar glueballs look like extra $\sigma$s and again it would be interesting to study a combined model of all these particles. This candidate is connected to the SM through operators of at least UV dimension 6 (such as $Tr F^2 h^2$) and so can have lifetimes longer than the age of the Universe. The glueballs could form part of the final dark matter mix with $\pi$. Again this mixture and its interactions remain to be explored and could provide interesting models. 

\section{Other Strongly Coupled Possibilities}
\label{sec:dm_discussion}

In this final section we note some other lightest states in strongly coupled theories which could serve as dark matter scenarios simply for completeness. We have sought to identify the lightest hadronic states in theories. Were these unstable in some model then it is possible that baryons could remain stable as the  dark matter relic - see for example~\cite{Kribs:2016cew,Cline:2021itd,Butterworth:2021jto}.

\subsection{Glueball Dark Matter in Pure Yang-Mills or Heavy Quark Theories}

At very large quark mass  the dominant scale for meson physics is $m_Q$ and all meson masses are approximately $2 m_Q$. The holographic model shows this behaviour at large quark masses but we do not show these results because the holographic approach assumes the gauge dynamics to be strongly coupled whereas the gauge theories become weakly coupled at the quark mass scale. Below the quark mass the gauge dynamics becomes a pure Yang-Mills theory. At large quark mass the glueball sector will be lightest part of the spectrum and these theories are not pionic dark matter theories.  Glueball dark matter is discussed in e.g.~\cite{daRocha:2017cxu,Carenza:2022pjd,Carenza:2023eua,Batz:2023zef,McKeen:2024trt,Soni:2016gzf}. 

\subsection{Unparticles}

Theories with $\nf > N_f^c$ and massless quarks live in the conformal window. The IR theory is a conformal plasma known as unparticles \cite{Georgi:2007ek}. Since the unparticle plasma is massless it can't be dark matter.  As soon as a mass is introduced one enters our region 5 (if the fixed point is strongly coupled) or region 9 if more weakly coupled. A first study of these later models can be found in \cite{Kikuchi:2007az}. We note here that although the mass hierarchy in these scenarios resembles one of the regions in fig.~\ref{fig:boxes_1} (left panel), their low energy effective theory may be considerably different.

\subsection{$\sigma$ Dark Matter}
An interesting question that emerges from the above theory space scan is whether it is ever possible within the $\nf, \nc$ space we consider to have a theory in which the $\sigma$ meson is the lightest particle and hence a dark matter candidate. In the theories so far discussed the $\sigma$ is always heavier than the $\pi$ moving to degenerate in the extreme walking case. We now briefly describe two models that could possibly achieve this goal though. 

As we have discussed above a $\sigma$ is not a natural dark matter candidate because naively one would expect the presence of UV dimension five operators such as $\bar{Q}Q|h|^2 $ that become $\sigma |h|^2$ in the IR. These likely lead to the $\sigma$ being unstable relative to the lifetime of the Universe unless both the $\sigma$ is light and the UV completion scale is very high as discussed above. It is possible though that some $Z_2$ symmetry might eliminate these terms in a more UV complete model (this is also true in the cases with a light $\sigma$ already discussed). We therefore include this discussion for completeness.
\subsubsection{$\nf=1 +N_X$ Theories}
One possible way to achieve the $\sigma$ as the lightest particle is to consider SU($\nc$) theories with $\nf=1$. Here the chiral symmetry is anomalous and the single ``pion" associated with the quark condensate will be heavy due to that anomaly~\cite{Farchioni:2007dw}. Our holographic model does not include the anomaly since it is based on dualities at large $\nc$ where the anomaly vanishes. The problem in this theory though is that $\nf=1$ theories are not walking and the $\sigma$ would be heavy. This could be repaired by including $N_X$ additional massive fermions with masses below but close to the dark sector strong coupling scale. All bound states containing the $X$ fermions would lie near the strong coupling or $X$ mass scale. The $\nf=1$ sector would experience the running dynamics of the $\nf=1+N_X$ so in the (anomalous) U(1) sector a light $\sigma$ would likely emerge made from the massless quark. 
\subsubsection{Chirally Gauged $\nf=2 +N_X$ Theories}
\begin{figure}
    \centering
    \includegraphics[width=0.5\linewidth]{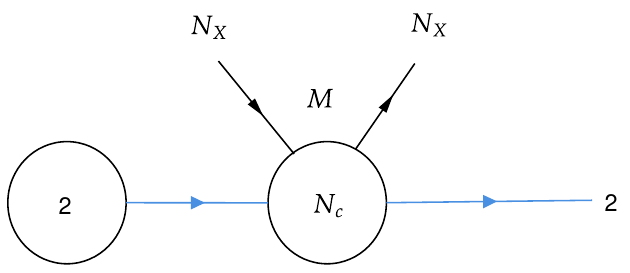}
    \caption{Matter content of the $N_f=2+N_X$ model as a moose diagram \cite{Chivukula:1987fw}. }
    \label{fig:moose}
\end{figure}

An alternative possibility is to remove the Nambu-Goldstone bosons from the light spectrum by having them eaten by a set of gauge bosons. Inspired by technicolour/the SM one could for example use $\nf=2$ and gauge the SU(2)$_{L_D}$ symmetry of the massless dark matter quarks (note this is not the SU(2)$_L$ of the SM). The matter content is shown in tab.~\ref{tab:matter_content} or as a moose  diagram \cite{Chivukula:1987fw} in Fig. \ref{fig:moose}. The massless pions will be eaten by the SU(2)$_{L_D}$ gauge fields and become part of a massive multiplet of gauge bosons with mass $g_2 \fpi$. Note the case where one gauges an SU(2)$_{L_D}$ sub-group, rather than some SU($\nf$) subgroup, is unique in that it has real representations and is anomaly free. That this dark matter sector mimics the SM structure is amusing! 

\begin{table}[h!]
\centering
\begin{tabular}{ccccc}
 & $SU(N_{c})$ & $SU(2)_{L_D}$ & $SU(2)_{R_D}$ & $SU(N_{X})_V$\\
 $q_L= \left( \begin{array}{c} u_L \\ d_L \end{array} \right)$  & $\Box$ & $\Box$ & 1 & 1\\
 $q_R^C = \left( \begin{array}{c} u_R^C \\ d_R^C \end{array} \right)$& $\overline{\Box}$ & 1 & $\overline{\Box}$ & 1 \\
 $\psi_{X,L}$ & $\Box$ & 1 & 1 & $\Box$ \\
 $\psi_{X,R}$ & $\overline{\Box}$ & 1 & 1 & $\Box$ \\
\end{tabular}
\caption{Matter representation for a possible $\sigma$ dark matter model.}
\label{tab:matter_content}
\end{table}

Again the problem is that the models with $\nf=2$ do not have walking dynamics to make the $\sigma$ light. As in the previous model we could solve this by including extra fermions with masses near the strong coupling scale to force a walking behaviour at the the chiral symmetry breaking scale. In fig.~\ref{fig:moose} we call these fermions $\psi_X$, they are a singlet of the SU(2)$_{L_D}$ gauge symmetry. It is likely a $\sigma$ made from the massless fermions will be the lightest state. 
\section{Conclusions}
\label{sec:conclusion}
Strongly-coupled theories provide interesting dark matter candidates in the form of pions, which produce relic density through number changing interactions within the strong-sector. Together with their ability to generate large self-interaction cross-section, they form a viable class of theories resulting in new phenomenology across cosmology and experimental searches. Due to their inherent non-perturbative nature these theories demand careful attention, especially to understand possible new relic density mechanisms beyond the well known ones. In this work, we embarked on such a survey using a holography model. Our aim was to demonstrate the lightest degrees of freedom in a large part of theory parameter space and elucidate associated potential dark matter phenomenology.

In this context, working within a SU($\nc = 3$) gauge groups with $\nf$ fermions in the fundamental representation, we sketched out seven distinct regions where qualitatively different dark matter phenomenology may appear. Regions where pions are the only relevant degrees of freedom are restricted and do not generate large enough $\mpi/\fpi$ to obtain phenomenologically consistent theory space. This points to analysis of additional relic density mechanisms adhering to the theory constraints. From this point of view, the next lightest species may play an important role.  An obvious example is $\rho, \pi$ only theories but they appear viable only in a very small region of parameter space within our investigations. A heavier quark mass or larger number of flavours, immediately leads to a light $\sigma$ which may have interesting consequences for dark matter phenomenology. Such $\rho, \pi,\sigma$ admixed regions have not yet been analysed in the literature to the best of our knowledge. This region is in fact the most dominant region in the parameter space we study. Complementary to these we also find regions where $\pi,\sigma$ can be simultaneously light, as one nears the conformal window.

Comparison of our results with non-perturbative calculations such as the lattice demonstrates that our mass spectrum is in  good agreement with first principles calculations. Our values of decay constant are less precise due to limitations of the holographic model itself, where large quark masses reveal strong coupling in the conformal UV that is nor present in the true gauge theories. Never-the-less, the qualitative understanding is correct. Variation with respect to $\nc$ for fixed $\nf/\nc$ is almost trivial, wherein no change in the spectrum is observed. This leads us to conclude that we can generalise our results across large regions of parameter space.

Finally we commented on other possible dark matter candidates such as the glueballs or unparticles with mass gap. We also briefly sketch confining theories where a light $\sigma$ can be obtained, keeping it as the only relevant degree of freedom in the theory. We argue that to stablise $\sigma$ against potential decays to the Standard Model, additional symmetry mechanisms may be necessary. 

Our investigation underlines the necessity of coupling non-perturbative understanding of strongly-interacting theories with dark matter phenomenology. Lattice studies while the best course of action for such an analysis, may not be computationally feasible. Therefore, use of holography techniques can serve as a first guiding principle in identifying interesting regions of parameter space as well as setting qualitative understanding of underlying parameters for dark matter. This understanding, followed by dark matter analysis will lead to useful feedback for the lattice community where targeted analyses may be carried out. We intend to embark on such a program in our follow up studies.

\section{Acknowledgements}
We thank Tom DeGrand and Ethan Neil for useful discussions.  S.K. is supported by the FWF research group funding FG1 and FWF project number P 36947-N. N.E.’s work was supported by the
STFC consolidated grant ST/X000583/1. 
W.P. is supported by the DFG project PO 1337/11-1. We are grateful to Mainz Institute for Theoretical Physics (MITP) of the Cluster of Excellence PRISMA$^+$ project (Project ID 390831469), for its hospitality and its partial support during this work. SK and NE thank W\"urzburg university theoretical particle physics group for hospitality during which part of the work was completed and the DFG founded RTG 2994 for corresponding support.


\bibliography{bibliography}

\providecommand{\href}[2]{#2}\begingroup\raggedright\begin{thebibliography}{10}

\bibitem{Hochberg:2014kqa}
Y.~Hochberg, E.~Kuflik, H.~Murayama, T.~Volansky and J.~G. Wacker,
  \textit{{Model for Thermal Relic Dark Matter of Strongly Interacting Massive
  Particles}},
  \href{https://doi.org/10.1103/PhysRevLett.115.021301}{\textit{Phys. Rev.
  Lett.} {\bfseries 115} (2015) 021301},
  [\href{https://arxiv.org/abs/1411.3727}{{\ttfamily 1411.3727}}].

\bibitem{Pomper:2024otb}
J.~Pomper and S.~Kulkarni, \textit{{Low energy effective theories of composite
  dark matter with real representations}},
  \href{https://arxiv.org/abs/2402.04176}{{\ttfamily 2402.04176}}.

\bibitem{Bernreuther:2019pfb}
E.~Bernreuther, F.~Kahlhoefer, M.~Kr\"amer and P.~Tunney, \textit{{Strongly
  interacting dark sectors in the early Universe and at the LHC through a
  simplified portal}},
  \href{https://doi.org/10.1007/JHEP01(2020)162}{\textit{JHEP} {\bfseries 01}
  (2020) 162}, [\href{https://arxiv.org/abs/1907.04346}{{\ttfamily
  1907.04346}}].

\bibitem{Bernreuther:2023kcg}
E.~Bernreuther, N.~Hemme, F.~Kahlhoefer and S.~Kulkarni, \textit{{Dark matter
  relic density in strongly interacting dark sectors with light vector
  mesons}}, \href{https://doi.org/10.1103/PhysRevD.110.035009}{\textit{Phys.
  Rev. D} {\bfseries 110} (2024) 035009},
  [\href{https://arxiv.org/abs/2311.17157}{{\ttfamily 2311.17157}}].

\bibitem{Choi:2018iit}
S.-M. Choi, H.~M. Lee, P.~Ko and A.~Natale, \textit{{Resolving phenomenological
  problems with strongly-interacting-massive-particle models with dark vector
  resonances}}, \href{https://doi.org/10.1103/PhysRevD.98.015034}{\textit{Phys.
  Rev. D} {\bfseries 98} (2018) 015034},
  [\href{https://arxiv.org/abs/1801.07726}{{\ttfamily 1801.07726}}].

\bibitem{Appelquist:2024koa}
T.~Appelquist, J.~Ingoldby and M.~Piai, \textit{{Dilaton Forbidden Dark
  Matter}},  \href{https://arxiv.org/abs/2404.07601}{{\ttfamily 2404.07601}}.

\bibitem{DelDebbio:2021xwu}
L.~Del~Debbio and R.~Zwicky, \textit{{Dilaton and massive hadrons in a
  conformal phase}},
  \href{https://doi.org/10.1007/JHEP08(2022)007}{\textit{JHEP} {\bfseries 08}
  (2022) 007}, [\href{https://arxiv.org/abs/2112.11363}{{\ttfamily
  2112.11363}}].

\bibitem{Zwicky:2023krx}
R.~Zwicky, \textit{{QCD with an infrared fixed point and a dilaton}},
  \href{https://doi.org/10.1103/PhysRevD.110.014048}{\textit{Phys. Rev. D}
  {\bfseries 110} (2024) 014048},
  [\href{https://arxiv.org/abs/2312.13761}{{\ttfamily 2312.13761}}].

\bibitem{Alho:2013dka}
T.~Alho, N.~Evans and K.~Tuominen, \textit{{Dynamic AdS/QCD and the Spectrum of
  Walking Gauge Theories}},
  \href{https://doi.org/10.1103/PhysRevD.88.105016}{\textit{Phys. Rev. D}
  {\bfseries 88} (2013) 105016},
  [\href{https://arxiv.org/abs/1307.4896}{{\ttfamily 1307.4896}}].

\bibitem{Erdmenger:2020flu}
J.~Erdmenger, N.~Evans, W.~Porod and K.~S. Rigatos, \textit{{Gauge/gravity dual
  dynamics for the strongly coupled sector of composite Higgs models}},
  \href{https://doi.org/10.1007/JHEP02(2021)058}{\textit{JHEP} {\bfseries 02}
  (2021) 058}, [\href{https://arxiv.org/abs/2010.10279}{{\ttfamily
  2010.10279}}].

\bibitem{Clemens:2017udk}
W.~Clemens and N.~Evans, \textit{{A Holographic Study of the Gauged NJL
  Model}}, \href{https://doi.org/10.1016/j.physletb.2017.05.027}{\textit{Phys.
  Lett. B} {\bfseries 771} (2017) 1--4},
  [\href{https://arxiv.org/abs/1702.08693}{{\ttfamily 1702.08693}}].

\bibitem{Evans:2013vca}
N.~Evans and K.~Tuominen, \textit{{Holographic modelling of a light
  technidilaton}},
  \href{https://doi.org/10.1103/PhysRevD.87.086003}{\textit{Phys. Rev. D}
  {\bfseries 87} (2013) 086003},
  [\href{https://arxiv.org/abs/1302.4553}{{\ttfamily 1302.4553}}].

\bibitem{Erdmenger:2020lvq}
J.~Erdmenger, N.~Evans, W.~Porod and K.~S. Rigatos, \textit{{Gauge/gravity
  dynamics for composite Higgs models and the top mass}},
  \href{https://doi.org/10.1103/PhysRevLett.126.071602}{\textit{Phys. Rev.
  Lett.} {\bfseries 126} (2021) 071602},
  [\href{https://arxiv.org/abs/2009.10737}{{\ttfamily 2009.10737}}].

\bibitem{Alfano:2024aek}
A.~Alfano and N.~Evans, \textit{{Mass hierarchies in gauge theory with
  two-index symmetric representation matter}},
  \href{https://doi.org/10.1103/PhysRevD.111.026001}{\textit{Phys. Rev. D}
  {\bfseries 111} (2025) 026001},
  [\href{https://arxiv.org/abs/2409.07977}{{\ttfamily 2409.07977}}].

\bibitem{Alfano:2025dch}
A.~Alfano, N.~Evans and W.~Fan, \textit{{Holography for QCD(Adj) and
  QCD(Adj)+F}},  \href{https://arxiv.org/abs/2506.09456}{{\ttfamily
  2506.09456}}.

\bibitem{daRocha:2017cxu}
R.~da~Rocha, \textit{{Dark SU(N) glueball stars on fluid branes}},
  \href{https://doi.org/10.1103/PhysRevD.95.124017}{\textit{Phys. Rev. D}
  {\bfseries 95} (2017) 124017},
  [\href{https://arxiv.org/abs/1701.00761}{{\ttfamily 1701.00761}}].

\bibitem{Carenza:2022pjd}
P.~Carenza, R.~Pasechnik, G.~Salinas and Z.-W. Wang, \textit{{Glueball Dark
  Matter Revisited}},
  \href{https://doi.org/10.1103/PhysRevLett.129.261302}{\textit{Phys. Rev.
  Lett.} {\bfseries 129} (2022) 261302},
  [\href{https://arxiv.org/abs/2207.13716}{{\ttfamily 2207.13716}}].

\bibitem{Carenza:2023eua}
P.~Carenza, T.~Ferreira, R.~Pasechnik and Z.-W. Wang, \textit{{Glueball dark
  matter}}, \href{https://doi.org/10.1103/PhysRevD.108.123027}{\textit{Phys.
  Rev. D} {\bfseries 108} (2023) 123027},
  [\href{https://arxiv.org/abs/2306.09510}{{\ttfamily 2306.09510}}].

\bibitem{Batz:2023zef}
A.~Batz, T.~Cohen, D.~Curtin, C.~Gemmell and G.~D. Kribs, \textit{{Dark sector
  glueballs at the LHC}},
  \href{https://doi.org/10.1007/JHEP04(2024)070}{\textit{JHEP} {\bfseries 04}
  (2024) 070}, [\href{https://arxiv.org/abs/2310.13731}{{\ttfamily
  2310.13731}}].

\bibitem{McKeen:2024trt}
D.~McKeen, R.~Mizuta, D.~E. Morrissey and M.~Shamma, \textit{{Dark matter from
  dark glueball dominance}},
  \href{https://doi.org/10.1103/PhysRevD.111.015044}{\textit{Phys. Rev. D}
  {\bfseries 111} (2025) 015044},
  [\href{https://arxiv.org/abs/2406.18635}{{\ttfamily 2406.18635}}].

\bibitem{Soni:2016gzf}
A.~Soni and Y.~Zhang, \textit{{Hidden SU(N) Glueball Dark Matter}},
  \href{https://doi.org/10.1103/PhysRevD.93.115025}{\textit{Phys. Rev. D}
  {\bfseries 93} (2016) 115025},
  [\href{https://arxiv.org/abs/1602.00714}{{\ttfamily 1602.00714}}].

\bibitem{Cline:2021itd}
J.~M. Cline, \textit{{Dark atoms and composite dark matter}},
  \href{https://doi.org/10.21468/SciPostPhysLectNotes.52}{\textit{SciPost Phys.
  Lect. Notes} {\bfseries 52} (2022) 1},
  [\href{https://arxiv.org/abs/2108.10314}{{\ttfamily 2108.10314}}].

\bibitem{Kribs:2016cew}
G.~D. Kribs and E.~T. Neil, \textit{{Review of strongly-coupled composite dark
  matter models and lattice simulations}},
  \href{https://doi.org/10.1142/S0217751X16430041}{\textit{Int. J. Mod. Phys.
  A} {\bfseries 31} (2016) 1643004},
  [\href{https://arxiv.org/abs/1604.04627}{{\ttfamily 1604.04627}}].

\bibitem{Georgi:2007ek}
H.~Georgi, \textit{{Unparticle physics}},
  \href{https://doi.org/10.1103/PhysRevLett.98.221601}{\textit{Phys. Rev.
  Lett.} {\bfseries 98} (2007) 221601},
  [\href{https://arxiv.org/abs/hep-ph/0703260}{{\ttfamily hep-ph/0703260}}].

\bibitem{Kikuchi:2007az}
T.~Kikuchi and N.~Okada, \textit{{Unparticle Dark Matter}},
  \href{https://doi.org/10.1016/j.physletb.2008.06.021}{\textit{Phys. Lett. B}
  {\bfseries 665} (2008) 186--189},
  [\href{https://arxiv.org/abs/0711.1506}{{\ttfamily 0711.1506}}].

\bibitem{Erlich:2005qh}
J.~Erlich, E.~Katz, D.~T. Son and M.~A. Stephanov, \textit{{QCD and a
  holographic model of hadrons}},
  \href{https://doi.org/10.1103/PhysRevLett.95.261602}{\textit{Phys. Rev.
  Lett.} {\bfseries 95} (2005) 261602},
  [\href{https://arxiv.org/abs/hep-ph/0501128}{{\ttfamily hep-ph/0501128}}].

\bibitem{DaRold:2005mxj}
L.~Da~Rold and A.~Pomarol, \textit{{Chiral symmetry breaking from five
  dimensional spaces}},
  \href{https://doi.org/10.1016/j.nuclphysb.2005.05.009}{\textit{Nucl. Phys. B}
  {\bfseries 721} (2005) 79--97},
  [\href{https://arxiv.org/abs/hep-ph/0501218}{{\ttfamily hep-ph/0501218}}].

\bibitem{Witten:1998qj}
E.~Witten, \textit{{Anti de Sitter space and holography}},
  \href{https://doi.org/10.4310/ATMP.1998.v2.n2.a2}{\textit{Adv. Theor. Math.
  Phys.} {\bfseries 2} (1998) 253--291},
  [\href{https://arxiv.org/abs/hep-th/9802150}{{\ttfamily hep-th/9802150}}].

\bibitem{Breitenlohner:1982jf}
P.~Breitenlohner and D.~Z. Freedman, \textit{{Stability in Gauged Extended
  Supergravity}},
  \href{https://doi.org/10.1016/0003-4916(82)90116-6}{\textit{Annals Phys.}
  {\bfseries 144} (1982) 249}.

\bibitem{Appelquist:1996dq}
T.~Appelquist, J.~Terning and L.~C.~R. Wijewardhana, \textit{{The Zero
  temperature chiral phase transition in SU(N) gauge theories}},
  \href{https://doi.org/10.1103/PhysRevLett.77.1214}{\textit{Phys. Rev. Lett.}
  {\bfseries 77} (1996) 1214--1217},
  [\href{https://arxiv.org/abs/hep-ph/9602385}{{\ttfamily hep-ph/9602385}}].

\bibitem{Hasenfratz:2023wbr}
A.~Hasenfratz, E.~T. Neil, Y.~Shamir, B.~Svetitsky and O.~Witzel,
  \textit{{Infrared fixed point of the SU(3) gauge theory with Nf=10 flavors}},
  \href{https://doi.org/10.1103/PhysRevD.108.L071503}{\textit{Phys. Rev. D}
  {\bfseries 108} (2023) L071503},
  [\href{https://arxiv.org/abs/2306.07236}{{\ttfamily 2306.07236}}].

\bibitem{Goertz:2024dnz}
F.~Goertz, {\'A}.~Pastor-Guti{\'e}rrez and J.~M. Pawlowski,
  \textit{{Gauge-Fermion Cartography: from confinement and chiral symmetry
  breaking to conformality}},
  \href{https://arxiv.org/abs/2412.12254}{{\ttfamily 2412.12254}}.

\bibitem{Veneziano:1968yb}
G.~Veneziano, \textit{{Construction of a crossing - symmetric, Regge behaved
  amplitude for linearly rising trajectories}},
  \href{https://doi.org/10.1007/BF02824451}{\textit{Nuovo Cim. A} {\bfseries
  57} (1968) 190--197}.

\bibitem{Albouy:2022cin}
G.~Albouy et~al., \textit{{Theory, phenomenology, and experimental avenues for
  dark showers: a Snowmass 2021 report}},
  \href{https://doi.org/10.1140/epjc/s10052-022-11048-8}{\textit{Eur. Phys. J.
  C} {\bfseries 82} (2022) 1132},
  [\href{https://arxiv.org/abs/2203.09503}{{\ttfamily 2203.09503}}].

\bibitem{DeGrand:2019vbx}
T.~DeGrand and E.~T. Neil, \textit{{Repurposing lattice QCD results for
  composite phenomenology}},
  \href{https://doi.org/10.1103/PhysRevD.101.034504}{\textit{Phys. Rev. D}
  {\bfseries 101} (2020) 034504},
  [\href{https://arxiv.org/abs/1910.08561}{{\ttfamily 1910.08561}}].

\bibitem{Bali:2013kia}
G.~S. Bali, F.~Bursa, L.~Castagnini, S.~Collins, L.~Del~Debbio, B.~Lucini
  et~al., \textit{{Mesons in large-N QCD}},
  \href{https://doi.org/10.1007/JHEP06(2013)071}{\textit{JHEP} {\bfseries 06}
  (2013) 071}, [\href{https://arxiv.org/abs/1304.4437}{{\ttfamily 1304.4437}}].

\bibitem{Bali:2008an}
G.~S. Bali and F.~Bursa, \textit{{Mesons at large N(c) from lattice QCD}},
  \href{https://doi.org/10.1088/1126-6708/2008/09/110}{\textit{JHEP} {\bfseries
  09} (2008) 110}, [\href{https://arxiv.org/abs/0806.2278}{{\ttfamily
  0806.2278}}].

\bibitem{DelDebbio:2007wk}
L.~Del~Debbio, B.~Lucini, A.~Patella and C.~Pica, \textit{{Quenched mesonic
  spectrum at large N}},
  \href{https://doi.org/10.1088/1126-6708/2008/03/062}{\textit{JHEP} {\bfseries
  03} (2008) 062}, [\href{https://arxiv.org/abs/0712.3036}{{\ttfamily
  0712.3036}}].

\bibitem{Fischer:2006ub}
C.~S. Fischer, \textit{{Infrared properties of QCD from Dyson-Schwinger
  equations}}, \href{https://doi.org/10.1088/0954-3899/32/8/R02}{\textit{J.
  Phys. G} {\bfseries 32} (2006) R253--R291},
  [\href{https://arxiv.org/abs/hep-ph/0605173}{{\ttfamily hep-ph/0605173}}].

\bibitem{Holdom:1981rm}
B.~Holdom, \textit{{Raising the Sideways Scale}},
  \href{https://doi.org/10.1103/PhysRevD.24.1441}{\textit{Phys. Rev. D}
  {\bfseries 24} (1981) 1441}.

\bibitem{Wittman:2017gxn}
D.~Wittman, N.~Golovich and W.~A. Dawson, \textit{{The Mismeasure of Mergers:
  Revised Limits on Self-interacting Dark Matter in Merging Galaxy Clusters}},
  \href{https://doi.org/10.3847/1538-4357/aaee77}{\textit{Astrophys. J.}
  {\bfseries 869} (2018) 104},
  [\href{https://arxiv.org/abs/1701.05877}{{\ttfamily 1701.05877}}].

\bibitem{Randall:2008ppe}
S.~W. Randall, M.~Markevitch, D.~Clowe, A.~H. Gonzalez and M.~Bradac,
  \textit{{Constraints on the Self-Interaction Cross-Section of Dark Matter
  from Numerical Simulations of the Merging Galaxy Cluster 1E 0657-56}},
  \href{https://doi.org/10.1086/587859}{\textit{Astrophys. J.} {\bfseries 679}
  (2008) 1173--1180}, [\href{https://arxiv.org/abs/0704.0261}{{\ttfamily
  0704.0261}}].

\bibitem{Robertson:2016xjh}
A.~Robertson, R.~Massey and V.~Eke, \textit{{What does the Bullet Cluster tell
  us about self-interacting dark matter?}},
  \href{https://doi.org/10.1093/mnras/stw2670}{\textit{Mon. Not. Roy. Astron.
  Soc.} {\bfseries 465} (2017) 569--587},
  [\href{https://arxiv.org/abs/1605.04307}{{\ttfamily 1605.04307}}].

\bibitem{Hansen:2015yaa}
M.~Hansen, K.~Lang\ae{}ble and F.~Sannino, \textit{{SIMP model at NNLO in
  chiral perturbation theory}},
  \href{https://doi.org/10.1103/PhysRevD.92.075036}{\textit{Phys. Rev. D}
  {\bfseries 92} (2015) 075036},
  [\href{https://arxiv.org/abs/1507.01590}{{\ttfamily 1507.01590}}].

\bibitem{Braat:2023fhn}
P.~Braat and M.~Postma, \textit{{SIMPly add a dark photon}},
  \href{https://doi.org/10.1007/JHEP03(2023)216}{\textit{JHEP} {\bfseries 03}
  (2023) 216}, [\href{https://arxiv.org/abs/2301.04513}{{\ttfamily
  2301.04513}}].

\bibitem{Chu:2024rrv}
X.~Chu, M.~Nikolic and J.~Pradler, \textit{{Even SIMP miracles are possible}},
  \href{https://arxiv.org/abs/2401.12283}{{\ttfamily 2401.12283}}.

\bibitem{Carmona:2024tkg}
A.~Carmona, F.~Elahi, C.~Scherb and P.~Schwaller, \textit{{Dark showers from
  sneaky dark matter}},
  \href{https://doi.org/10.1007/JHEP06(2025)198}{\textit{JHEP} {\bfseries 06}
  (2025) 198}, [\href{https://arxiv.org/abs/2411.15073}{{\ttfamily
  2411.15073}}].

\bibitem{Alexander:2023wgk}
S.~Alexander, H.~Gilmer, T.~Manton and E.~McDonough,
  \textit{{{\ensuremath{\pi}}-axion and {\ensuremath{\pi}}-axiverse of dark
  QCD}}, \href{https://doi.org/10.1103/PhysRevD.108.123014}{\textit{Phys. Rev.
  D} {\bfseries 108} (2023) 123014},
  [\href{https://arxiv.org/abs/2304.11176}{{\ttfamily 2304.11176}}].

\bibitem{Davighi:2024zip}
J.~Davighi, A.~Greljo and N.~Selimovic, \textit{{Topological Portal to the Dark
  Sector}}, \href{https://doi.org/10.1103/PhysRevLett.134.111804}{\textit{Phys.
  Rev. Lett.} {\bfseries 134} (2025) 111804},
  [\href{https://arxiv.org/abs/2401.09528}{{\ttfamily 2401.09528}}].

\bibitem{Davighi:2025awm}
J.~Davighi, S.~Moldovsky, H.~Murayama, C.~Scherb and N.~Selimovic,
  \textit{{Topological Freeze-out by Semi-Annihilation}},
  \href{https://arxiv.org/abs/2506.05468}{{\ttfamily 2506.05468}}.

\bibitem{Hong:2019nwd}
S.~Hong, G.~Kurup and M.~Perelstein, \textit{{Conformal Freeze-In of Dark
  Matter}}, \href{https://doi.org/10.1103/PhysRevD.101.095037}{\textit{Phys.
  Rev. D} {\bfseries 101} (2020) 095037},
  [\href{https://arxiv.org/abs/1910.10160}{{\ttfamily 1910.10160}}].

\bibitem{Hong:2022gzo}
S.~Hong, G.~Kurup and M.~Perelstein, \textit{{Dark matter from a conformal Dark
  Sector}}, \href{https://doi.org/10.1007/JHEP02(2023)221}{\textit{JHEP}
  {\bfseries 02} (2023) 221},
  [\href{https://arxiv.org/abs/2207.10093}{{\ttfamily 2207.10093}}].

\bibitem{Cho:2007cy}
Y.~M. Cho and J.~H. Kim, \textit{{Dilatonic dark matter and its experimental
  detection}}, \href{https://doi.org/10.1103/PhysRevD.79.023504}{\textit{Phys.
  Rev. D} {\bfseries 79} (2009) 023504},
  [\href{https://arxiv.org/abs/0711.2858}{{\ttfamily 0711.2858}}].

\bibitem{Cho:1998aa}
Y.~M. Cho and Y.~Y. Keum, \textit{{Dilatonic dark matter: A new paradigm}},
  \href{https://doi.org/10.1142/S0217732398000152}{\textit{Mod. Phys. Lett. A}
  {\bfseries 13} (1998) 109--117},
  [\href{https://arxiv.org/abs/hep-ph/9810379}{{\ttfamily hep-ph/9810379}}].

\bibitem{Kondo:2022lgg}
D.~Kondo, R.~McGehee, T.~Melia and H.~Murayama, \textit{{Linear sigma dark
  matter}}, \href{https://doi.org/10.1007/JHEP09(2022)041}{\textit{JHEP}
  {\bfseries 09} (2022) 041},
  [\href{https://arxiv.org/abs/2205.08088}{{\ttfamily 2205.08088}}].

\bibitem{Farchioni:2007dw}
F.~Farchioni, I.~Montvay, G.~Munster, E.~E. Scholz, T.~Sudmann and J.~Wuilloud,
  \textit{{Hadron masses in QCD with one quark flavour}},
  \href{https://doi.org/10.1140/epjc/s10052-007-0394-4}{\textit{Eur. Phys. J.
  C} {\bfseries 52} (2007) 305--314},
  [\href{https://arxiv.org/abs/0706.1131}{{\ttfamily 0706.1131}}].

\bibitem{Butterworth:2021jto}
J.~M. Butterworth, L.~Corpe, X.~Kong, S.~Kulkarni and M.~Thomas, \textit{{New
  sensitivity of LHC measurements to composite dark matter models}},
  \href{https://doi.org/10.1103/PhysRevD.105.015008}{\textit{Phys. Rev. D}
  {\bfseries 105} (2022) 015008},
  [\href{https://arxiv.org/abs/2105.08494}{{\ttfamily 2105.08494}}].

\bibitem{Chivukula:1987fw}
R.~S. Chivukula, H.~Georgi and L.~Randall, \textit{{A Composite Technicolor
  Standard Model of Quarks}},
  \href{https://doi.org/10.1016/0550-3213(87)90638-9}{\textit{Nucl. Phys. B}
  {\bfseries 292} (1987) 93--108}.

\end{thebibliography}\endgroup
\end{document}